\documentclass[superscriptaddress,aps,preprintnumbers,amsmath,showpacs,amssymb,prd,nofootinbib,reprint]{revtex4-1}
\usepackage{bm,color} 
\usepackage{amssymb,amsfonts,slashed,amsthm,amsmath,graphicx, soul}
\usepackage{epsfig}
\usepackage{ulem}

\begin{document}

\renewcommand{\figurename}{Fig.}
\renewcommand{\tablename}{Table.}
\newcommand{\Slash}[1]{{\ooalign{\hfil#1\hfil\crcr\raise.167ex\hbox{/}}}}
\newcommand{\bra}[1]{ \langle {#1} | }
\newcommand{\ket}[1]{ | {#1} \rangle }
\newcommand{\beq}{\begin{equation}}  \newcommand{\eeq}{\end{equation}}
\newcommand{\bef}{\begin{figure}}  \newcommand{\eef}{\end{figure}}
\newcommand{\bec}{\begin{center}}  \newcommand{\eec}{\end{center}}
\newcommand{\non}{\nonumber}  \newcommand{\eqn}[1]{\begin{equation} {#1}\end{equation}}
\newcommand{\laq}[1]{\label{eq:#1}}  
\newcommand{\dd}[1]{{d \o d{#1}}}
\newcommand{\Eq}[1]{Eq.~(\ref{eq:#1})}
\newcommand{\Eqs}[1]{Eqs.~(\ref{eq:#1})}
\newcommand{\eq}[1]{(\ref{eq:#1})}
\newcommand{\Sec}[1]{Sec.\ref{chap:#1}}
\newcommand{\ab}[1]{\left|{#1}\right|}
\newcommand{\vev}[1]{ \left\langle {#1} \right\rangle }
\newcommand{\bs}[1]{ {\boldsymbol {#1}} }
\newcommand{\lac}[1]{\label{chap:#1}}
\newcommand{\SU}[1]{{\rm SU{#1} } }
\newcommand{\SO}[1]{{\rm SO{#1}} }
\def\({\left(}
\def\){\right)}
\def\dt{{d \o dt}}
\def\diag{\mathop{\rm diag}\nolimits}
\def\Spin{\mathop{\rm Spin}}
\def\O{\mathcal{O}}
\def\U{\mathop{\rm U}}
\def\Sp{\mathop{\rm Sp}}
\def\SL{\mathop{\rm SL}}
\def\tr{\mathop{\rm tr}}
\def\ebq{\end{equation} \begin{equation}}
\newcommand{\OR}{~{\rm or}~}
\newcommand{\AND}{~{\rm and}~}
\newcommand{\EV}{ {\rm \, eV} }
\newcommand{\KEV}{ {\rm \, keV} }
\newcommand{\MEV}{ {\rm \, MeV} }
\newcommand{\GEV}{ {\rm \, GeV} }
\newcommand{\TEV}{ {\rm \, TeV} }
\def\o{\over}
\def\a{\alpha}
\def\b{\beta}
\def\c{\varepsilon}
\def\d{\delta}
\def\e{\epsilon}
\def\f{\phi}
\def\g{\gamma}
\def\h{\theta}
\def\k{\kappa}
\def\l{\lambda}
\def\m{\mu}
\def\n{\nu}
\def\p{\psi}
\def\q{\partial}
\def\r{\rho}
\def\s{\sigma}
\def\t{\tau}
\def\u{\upsilon}
\def\w{\omega}
\def\x{\xi}
\def\y{\eta}
\def\z{\zeta}
\def\D{\Delta}
\def\G{\Gamma}
\def\H{\Theta}
\def\L{\Lambda}
\def\F{\Phi}
\def\P{\Psi}
\def\S{\Sigma}
\def\me{\mathrm e}
\def\ol{\overline}
\def\tl{\tilde}
\def\*{\dagger}
\def\red#1{\textcolor{red}{#1}}
\def\WY#1{\textcolor{blue}{#1}}
\def\WYC#1{\textcolor{blue}{\bf { WY:} #1}}
\def\WYS#1{\textcolor{blue}{\sout{#1}}}

%%%%%%%%%%%%%%%%%%%%%%%%%%%%%%%%%%%%%%%%%%%%%%%%%%%%%%%%%%%%%%%

%######################
\preprint{TU-1177}
%######################

\title{
Thermal production of cold ``hot dark matter" around eV}

\author{
Wen Yin
}
\affiliation{Department of Physics, Tohoku University, 
Sendai, Miyagi 980-8578, Japan}

\begin{abstract}
A very simple production mechanism of feebly interacting dark matter (DM) that rarely annihilates is thermal production, which predicts the DM mass around eV. This has been widely known as the hot DM scenario. Despite there are several observational hints from background lights suggesting a DM in this mass range, the hot DM scenario has been considered strongly in tension with the structure formation of our Universe because the free-streaming length of the DM produced from thermal reactions was thought to be too long. 
 In this paper, I show that the previous conclusions are not always true depending on the reaction for bosonic DM because of the Bose-enhanced reaction at very low momentum.  
By using the simple $1\leftrightarrow 2$ decay/inverse decay process to produce the DM, I demonstrate that the eV range bosonic DM can be thermally produced {\it coldly} from a hot plasma by performing a model-independent analysis applicable to axion, hidden photon, and other bosonic DM candidates.
Therefore, the bosonic DM in the eV mass range may still be special and theoretically well-motivated. 
\end{abstract}

\maketitle
\flushbottom

\vspace{1cm}

%%%%%%%%%%%%%%%%%%%%%%%%%%%%%%%%%%%%%%%%%%%%%%%%%%%%%%%%%%%%%%%%%
\section{Introduction}
The origin of the dark matter (DM) of our Universe has been one of the leading mysteries of particle theory, cosmology, and astronomy for around a century~\cite{Planck:2018vyg}. 
A few decades ago, thermally produced feebly-interacting DM in the eV mass range was popularly considered, with a leading candidate of the standard model (SM) neutrino~(see, e.g., \cite{Frenk:2012ph}). Indeed, if the feebly-interacting DM once reaches the thermal equilibrium with the thermal bath in the early Universe, the number density of the DM is around that of the photon relic, and the matter-radiation equality, which is known to be around eV temperature, happens at the cosmic temperature around the DM mass. Then the DM mass is predicted around eV. 
This scenario is well known as the hot DM, which has been considered highly in tension with the structure formation. To be consistent, we need an entropy dilution making DM heavier than a few keV~\cite{Viel:2005qj, Irsic:2017ixq}.  Moreover, if the DM is a fermion, the Pauli exclusion principle for the DM in galaxies, i.e., the Tremaine-Gunn bound, excludes the mass below $\sim 100\EV$~\cite{Tremaine:1979we,Boyarsky:2008ju,Randall:2016bqw}. 
If the DM is not fully thermalized in the early Universe, e.g., it is produced from a freeze-in mechanism~\cite{Hall:2009bx}, the free-streaming bound still restricts the mass above several keVs and excludes the eV mass range~\cite{Kamada:2019kpe, DEramo:2020gpr}. In any case, DM interacting with thermal plasma na\"{i}vely acquires momentum around the cosmic temperature, and if the DM is lighter than a keV, the free-streaming length will be too long. 
It seems that any DM from thermal production with eV mass range is a no-go.  % is more free-parameter. 
On the other hand, recently, there have been hints from the observations of anisotropic cosmic infrared background and TeV gamma-ray spectrum, independently suggesting an axion-like particle (ALP) DM around the eV mass range 
 \cite{Gong:2015hke, Korochkin:2019qpe,  Caputo:2020msf, Bernal:2022xyi} (see also \cite{Kohri:2017oqn, Kalashev:2018bra, Kashlinsky:2018mnu}).\footnote{In contrast, the anisotropic cosmic infrared background data and the TeV gamma-ray spectrum suggest that the LORRI excess \cite{Lauer:2022fgc} cannot be simply explained by the decay of cold DM~\cite{Nakayama:2022jza,Carenza:2023qxh}(See also \cite{Bernal:2022wsu, Bernal:2022xyi}).} 
 In the future, there will be various experiments confirming the eV range DM, like the direct detection~\cite{Baryakhtar:2018doz},  indirect detection~\cite{Bessho:2022yyu}, line-intensity mapping~\cite{Shirasaki:2021yrp} (see also some experiments for a generic ALP including this mass range~, e.g., solar axion helioscope \cite{ Irastorza:2011gs, Armengaud:2014gea, Armengaud:2019uso, Abeln:2020ywv} and photon collider~\cite{Homma:2022ktv}). 
In this paper, I study if the aforementioned no-go theorem for the eV range DM is true. I will show by using a concrete example that the {\it cold} eV-range bosonic DM can be produced via the thermal interaction with hot plasma by taking into account the Bose-enhancement effect. 
% his may be a good period to study carefully how the eV DM can be produced.  

As we mentioned, the DM, much lighter than keV, is very likely to be a bosonic one due to the Tremaine-Gunn bound. A known successful scenario that predicts the eV DM is the ALP miracle scenario~\cite{Daido:2017wwb, Daido:2017tbr}, where the ALP DM is also the inflaton, driving the cosmic inflation.  The potential of the ALP is assumed to have an upside-down symmetry, via which 
the mass, as well as the self-couplings, of the ALP in the vacuum, is related to that during the hilltop inflation. The DM is a remnant of inflaton from a predicted incomplete reheating.
Interestingly, the eV mass range is predicted from the conditions for explaining the DM abundance, and the cosmic-microwave background normalization and spectral index for the power spectrum of the scalar density perturbation.\footnote{Alternatively, there are also various simple DM production mechanisms in standard cosmology that are consistent (but not predict) the eV mass range, like the DM production via inflationary fluctuation~\cite{Graham:2018jyp,Guth:2018hsa, Ho:2019ayl, Graham:2015rva,Ema:2019yrd}, 
the light DM production via inflaton decay~\cite{Moroi:2020has, Moroi:2020bkq}, 
for ALP with modified potentials~\cite{Arias:2012az,Nakagawa:2020eeg,Marsh:2019bjr}. 
} 

In this paper, we study another simple production mechanism, predicting the eV mass range: the thermal production that was thought to be excluded in the early studies.
I show by considering a two-body decay/inverse decay process, 
\beq \laq{decay}\chi_1\leftrightarrow \chi_2 \f\eeq 
 of a thermal distributed mother particle, $\chi_1,$
 with a mass, $M_1 (\ll T)$, into two daughter particles, $\chi_2 \AND \f$, including a light bosonic particle, $\f$, 
has a burst population era of the low-momentum mode $p_{\f}^{\rm busrt}\sim M_1^2/T$ of $\f$. 
Here, $T$ is the cosmic temperature at which $\chi_1$ is thermalized.  
The burst production of $\f$ is triggered by the reaction in a timescale $\D t_{\rm ignition}\sim \(\frac{T^3}{M_1^3} \Gamma^{\rm rest}_{\chi_1\to \chi_2 \f}\)^{-1}$ with 
$\Gamma^{\rm rest}_{\chi_1\to \chi_2 \f}$ being the proper decay rate of $\chi_1 \to \chi_2 \f$. Immediately, the Bose-enhanced production of $\f$ populates the momentum modes around $p_\f^{\rm burst}$ until the $\f$ number density reaches about the number density of $\chi_1$. 
Thus low momentum modes of $\f$ are produced with a number density around $T^3.$ 
In the expanding Universe with the Hubble parameter, $H$, the condition for this to happen is $\D t_{\rm ignition}^{-1}\gg H$. 
The momentum of the cold component of $\f$ redshifts to be below $p_{\f}^{\rm burst}$ which blue shifts in time. 
If the usual thermalization rate of $\chi_2$ or $\f$, $\D t_{\rm decay}^{-1} \sim \G_{\chi_1\to \chi_2\f}^{\rm rest} M_1/T,$ is smaller than $H$ at the burst production, it cannot interact with $\chi_1,\chi_2$ anymore through \Eq{decay} due to kinematics, and the burst-produced $\f$ free-streams until it becomes non-relativistic.  Thus if the mass of $\f$ is around eV, we get the cold component abundance of $\f$ consistent with the measured DM abundance. 
Since the condition mostly relies on kinematics and statistics, this mechanism easily applies to produce generic bosonic DM, such as axion, hidden photon, and CP-even scalar (a candidate is CP-even ALP~\cite{Sakurai:2021ipp, Sakurai:2022cki, Haghighat:2022qyh} with dark sector PQ fermions~\cite{Sakurai:2021ipp}), etc.

The main difference from the previous approaches of freeze-in or thermal production of heavier DM is that I use the unintegrated Boltzmann equation for the evolution of the distribution functions of $\f$ and $\chi_2$ by including Bose-enhancement and Pauli-blocking effect as well as the mother particle mass effect. 
The important assumption is that $\chi_2,\f$ are both not thermalized when the burst production happens since I focus on light DM, which is typically considered to have a feeble interaction. 

The rest of this paper is organized as follows. In the next \Sec{2}, I will review the Boltzmann equation.  
In \Sec{3}, I use a simplified setup neglecting the expansion of the Universe to explain analytically and numerically my mechanism, the burst production of $\f$. 
In \Sec{4}, I will remove several assumptions made in \Sec{3}, and apply the mechanism to the DM production. 
I also discuss the conditions that the mechanism is not spoiled by other effects in more generic setups.
The last section \Sec{5} is devoted to discussion and conclusions, in which I will also comment on the application of the mechanism to produce the DM around keV.  

\section{Boltzmann equation}

\lac{2}
Let us study the production of $\f$ via \eq{decay}  by employing the standard (unintegrated) Boltzmann equation in expanding Universe  (see e.g.~\cite{Kolb:1990vq}), 
\begin{align}
\frac{\partial f_i[p_i, t]}{\partial t} - p_i H \frac{\partial f_i[p_i, t]}{\partial p_i} & = C^i[p_i, t],
% \\
%\frac{\partial f_\psi[p_\psi, t]}{\partial t} - p_\psi H \frac{\partial f_\psi[p_\psi, t]}{\partial p_\psi} & = C^\psi[p_\psi, t],
\end{align}
with $i=\chi_1,\chi_2, \f$ and $C^{i}$ being the collision term of $i$; 
$f_i$ is the distribution function of $i$; $H=\dot a/a$ is the Hubble parameter with $a$ being the scale factor; 
As aforementioned, I do not specify whether $\f$ is a vector or a scalar field (or a more generic field with integer spins) but I only assume that $\f$ is a boson while $\chi_{1,2}$ may be either fermions or bosons; I assumed the rotational invariance for the equations, and $p_i=|\vec{p}_i|$.

The collision term for, e.g., $i=\f$ of the $1\leftrightarrow 2$ process \eq{decay} has the form of 
\begin{align}
&\nonumber C^\f=   \frac{1}{ 2E_{\f }g_\f}  \sum \int \! d\Pi_{\chi_1} \, d\Pi_{\chi_2} \,\\
%\frac{d^3 p_1}{(2\pi)^3 2 E_1}   \frac{d^3 p_2}{(2\pi)^3 2 E_2} 
&\nonumber (2\pi)^4 \delta^4(p_{\chi_1}-p_{\phi}-p_{\chi_2})  
\laq{collision} \times |{\cal M}_{\chi_1\to \chi_2\f}|^2\\ &\times  S\(f_{\chi_1} [p_{\chi_1}], f_{\chi_2}[p_{\chi_2}], f_\f[p_{\f}]\)
%\left\{  f_{\chi_1}[p_{\chi_1}] (1-  f_{\chi_2}[p_{\chi_2}] ) ( 1+ f_{\f}[p_{\f}] ) - (1-  f_{\chi_1}[p_{\chi_1}] )   f_\phi[p_{\phi}]   f_{\chi_2}[p_{\chi_2}]   \right\}
. 
\end{align}
with $g_i$ being the internal degrees of freedom including spins, ${\cal M}_{\chi_1\to \chi_2\f}$ the amplitude, $d\Pi_{i}= \frac{d^3p_i}{2E_i (2\pi)^3}$ is phase space integral, the sum is performed over all internal degrees of freedom of the initial and final states,  
\begin{align}  S
&\equiv  f_{\chi_1}[p_{\chi_1}] (1\pm f_{\chi_2}[p_{\chi_2}] ) ( 1+ f_{\f}[p_{\f}] )\non\\ 
& ~~~- (1\pm f_{\chi_1}[p_{\chi_1}] )f_\phi[p_{\phi}]  f_{\chi_2}[p_{\chi_2}] \laq{S}
\\
&= \{{f_{\chi_1}(p_{\chi_1}) ( \pm  f_{\chi_2}(p_{\chi_2})+f_{\f}(p_{\f})+1)-f_{\chi_2}(p_{\chi_2})f_{\f}(p_{\f})}\}
 \end{align}
includes the Bose-enhancement and Pauli-exclusion effects. $+$ and $-$ correspond to the cases that $\chi_{1,2}$ are bosons and fermions, respectively.

\paragraph{Simplified form with only \eq{decay} reaction}
In this paper, I treat the reaction by \Eq{decay} seriously and make some approximations for the other possible reactions. 
%Thus, I consider the non-linear Boltzmann equation solely driven by \Eq{decay} the reaction.
By using the comoving momentum $\hat{p}_i=p_i a$, the Boltzmann equation with collision term \eq{collision} reduces to the simplified form 
\beq
\frac{d \hat f_\f[\hat{p}_\f]  \frac{\hat{p}_\f^2}{2\pi^2} }{dt} =  \frac{g_{\chi_1}}{g_\f} \int_{\hat p_{\chi_1}^-}^{\hat p_{\chi_1}^+} d \hat p_{\chi_1} \frac{\hat p_{\chi_1}^2}{2\pi^2}\hat f_{\chi_1}(\hat{p}_{\chi_1})\frac{\partial \Gamma_{\chi_1\to \chi_2 \f}^{\rm rest}}{\gamma\partial {\hat{ p}_\f}}  \frac{S[\hat{f}_i]}{\hat f_{\chi_1}(\hat{p}_{\chi_1})}.\laq{fphi}
\eeq
Here $\hat{p}^{\pm}_{{\chi}_1}$ and $d\Gamma^{\rm rest}_{\chi_1\to \chi_2 \f}/d\hat{p}_\f$ depend on the kinematics, which will be explained later, $\g=E_{\chi_1}/M_1$ (without a hat) the Lorentz factor, and $\hat{f}_i(\hat p_i)\equiv f_i(p_i).$ 
 This equation can be understood by multiplying $d \hat{p}_\f$ on both sides. Then the number density of $\phi$ in the momentum range $\hat{p}_\f \sim  \hat p_\f+ d \hat{p}_\f$ is produced by the decays minus inverse decays of $\chi_1$ in the whole kinematically-allowed phase space. The reaction rate is accompanied by the Lorentz factor, and the Bose-enhancement and Pauli-blocking factors, $S/\hat f_{\chi_1},$ which also includes the inverse decay effect.

The equations for the other particles can be similarly obtained, e.g., for $\chi_2$, we have
\beq
\frac{d f_{\chi_2}[\hat{p}_{\chi_2}]  \frac{\hat{p}_{\chi_2}^2}{2\pi^2} }{dt} =  \frac{g_{\chi_1}}{g_{\chi_2}} \int_{\hat p_{\chi_1}^-}^{\hat p_{\chi_1}^+} d \hat p_{\chi_1} \frac{\hat p_{\chi_1}^2}{2\pi^2}f_{\chi_1}(\hat{p}_{\chi_1})\frac{\partial \Gamma_{\chi_1\to \chi_2 \f}^{\rm rest}}{\g \partial {\hat{ p}_{\chi_2}}} \frac{S[\hat f_i]}{f_{\chi_1}(\hat{p}_{\chi_1})}.\laq{fpsi}
\eeq

The collision term via \eq{decay} conserves the difference of the comoving number density of $\chi_2$ minus that of $\f$, \beq 
\laq{conserv} (n_{\chi_2}-n_\f) a^3 =\rm const,\eeq 
as long as we do not have other fast interactions to change the comoving number of $\f$ or $\chi_2$. 
We also have $-\dt (n_{\chi_1} a^3)=\dt (n_{\f} a^3)=\dt (n_{\chi_2} a^3)$ in the case $\chi_1$ does not have other fast interaction than \eq{decay}. %I will mostly focus on \Eq{conserv}, since I will assume $\chi_1$ in the thermal equilibrium unless otherwise stated. 

\paragraph{Kinematics}
To discuss kinematics, let us first estimate the energy distribution in the boosted frame of $\chi_1$ moving along the z-axis with the Lorentz factor $\g=E_{\chi_1}/M_1$. 
The momentum of the injected $\f$, which has the momentum $p_\f^{\rm rest} = \frac{M_1^2-M_2^2}{M_1^2}\frac{M_1}{2}$ with an angle $\theta_{\chi_1}$ to the z-axis in the rest frame, is boosted as well 
\beq
\laq{mom}
{p}_{\f}= \(\gamma +\gamma \beta \cos{\theta_{\chi_1}} \) \times \eta \frac{M_{1}}{2}=\frac{\eta}{2}\(E_{\chi_1}+ p_{\chi_1} \cos \theta_{\chi_1}\)
\eeq
where $\beta =\sqrt{\g^2- 1}/\gamma$. 
%Although I take the mass  of $\chi_2$, $M_2$, $M_2=0$ in most of the simulations, 
%in the analytical formulas, 
I include the effect of the mass of $\chi_2$, $M_2$,  in \beq \eta \equiv \frac{M_1^2-M^2_2}{M^2_1}\eeq for generality and later convenience.  
However, I neglect the small $\f$ mass, $m_\f$, which is only taken into account when we estimate the DM energy density in the next section. 
Note that the lowest value of $p_\f$ is $\eta M_1^2/4p_{\chi_1}$ when $p_{\chi_1}\gg M_1,$ and $\cos \theta_{\chi_1}=-1$, i.e., $\f$ is injected  backward. 

Then I get 
\beq
\frac{\partial \Gamma^{\rm rest}_{\chi_1\to \chi_2 \f}}{\partial \hat p_\f}=  \frac{\Gamma^{\rm rest}_{\chi_1\to \chi_2 \f}}{\eta \hat p_{\chi_1}}.
\eeq
 Due to the rotational invariance, we do not have any preferred direction in the rest frame. 
$p_{\chi_1}^\pm$ can be obtained from \Eq{mom} with $\cos\theta_{\chi_1}$ in the range $[-1,1]$. \\

Before ending this section, I would like to note that given $\G_{\chi_1\to \chi_2 \f}^{\rm rest}$ $M_1, \eta, g_i$ and statistics, the equations are irrelevant to intrinsic interactions. This is the reason why I did not specify a model by using an explicit Lagrangian so far (for one explicit  Lagrangian, see \Sec{model}). 
In other words, the mechanism explained by the equations in this section should apply to large classes of models in which $\f$ is a bosonic particle.

\section{A burst production of bosonic particle}
\lac{3}
In this section, I will study the particle production of $\f$ carefully by using the Boltzmann equation in flat space introduced in the previous section. 
 For clarity, I use a simplified setup to describe the mechanism, and I will remove and discuss the simplifications in the next section, where the mechanism is applied to DM production in cosmology. 
The simplifications are listed as follows, 
\begin{description}
\item[Flat Universe]  I will neglect the expansion of the Universe, i.e., $a=1$, and I use \Eqs{fphi} and \eq{fpsi} by removing the hat. I will recover the effects of the expanding Universe in the following section. 
\item[Hierarchical timescales of other interactions] 
I assume for simplicity that $\chi_1$ is in the thermal equilibrium with the Bose-Einstein (Fermi-Dirac) distribution 
\beq
\laq{initial}
f_{\chi_1}\approx f_{\chi_1}^{\rm eq}\equiv\(e^{E_{\chi_1}/T}\mp 1 \)^{-1},
\eeq 
where $-\AND +$ are for the bosonic and fermionic $\chi_{1,2}$, respectively. Here and hereafter, I use the superscript ``eq" to denote the quantity in the thermal equilibrium. 
$f_{\chi_2}, \AND f_\f$ are treated as variables that evolve via \Eqs{fphi} and \eq{fpsi}.
This is a realistic condition if the reaction timescale for the other interactions for $\chi_1$ ($\chi_2, \f$) is much faster (slower) than 
the reactions induced by the process \eq{decay}. 
The fast reaction is assumed to keep $\chi_1$ always in the thermal equilibrium. 
I will come back to argue the case this is not satisfied in \Sec{slowthe}. 
\item[Initial conditions]
The initial conditions are taken as \beq f_{\chi_2,\f}[p_i]=0~~\text{at} ~~t=t_i \laq{tini} \eeq  for any $p_i$. 
I will comment on what happens by other initial conditions at the end of \Sec{thermalization}. 
\item[Relativistic plasma] 
I will focus on the case \beq T\gg M_1 \neq 0.\eeq 
%This is an important assumption for $\f$ production at low momentum. It is also important that the effect of $M_1$ is not neglected. 
In \Sec{4}, this assumption is removed in the expansion Universe.

\end{description}

\subsection{First stage: Ignition} 

Now we are ready to discuss particle production.
Let us focus on the mode of  $p_{\f}\ll M_1$. Then we get from \Eq{mom}, 
\beq
\laq{IRrange}
p_{\chi_1}^{-}\approx  \eta \frac{ M_1^2}{4p_\f},~~p_{\chi_1}^+= \infty
\eeq
Thus in \Eqs{fphi} and \eq{fpsi} only the higher momentum modes of $\chi_1$ can produce the lower momentum mode of $p_\f.$ 
In particular, by noting that the dominant mode of the thermal distributed $\chi_1$ has $p_{\chi_1}\sim  T (\gg M_1),$ $f_\f(p_\f)$ (not $f_\f p_\f^2 $) with momentum 
\beq 
\laq{pburst}
 p_\f \sim p_{\f}^{\rm burst} \equiv \eta \frac{M_1^2}{2T}
\eeq
is popularly produced. From the energy-momentum conservation with $p_\f^{\rm burst}\ll T$, \beq p_{\chi_1}, p_{\chi_2}\sim T\eeq  
in the reaction.
%In explaining the mechanism, I will mostly discuss the modes for $p_\f\sim p_\f^{\rm burst}$ and $E^{\rm burst}_{\chi_2}\sim E^{\rm burst}_{\chi_1}\sim T$ and I will also discuss the periods that the other modes are important in \Sec{thermalization}.
This first stage is characterized by conditions close to the initial one, 
\beq
\laq{stage1}
f_\f[p_\f\sim p_\f^{\rm burst}] \lesssim 1 \AND f_{\chi_2}[p_{\chi_2}\sim T]\ll 1,
\eeq
and the other momentum modes are also suppressed. 
Let us follow the evolution of $f_\f [p_\f\sim p_\f^{\rm burst}].$ 
By noting $ S/f_{\chi_1} \simeq 1$, a timescale that $f_\f (p_\f\sim p_\f^{\rm burst})$ reaches unity is derived from  \Eqs{fphi}, \eq{fpsi} and \eq{stage1} as
\beq
\D t_{\rm ignition}^{-1} %\sim \Gamma_{\rm trigger}
\sim  \frac{g_{\chi_1}}{g_\f} \frac{4T^3}{\eta^3 M_1^3} \Gamma^{\rm rest}_{\chi_1\to \chi_2 \f}.
\eeq
We note that at this timescale, $\chi_1$ rarely decays because the thermally averaged decay rate is 
 \beq \D t_{\rm decay}^{-1}\sim \Gamma^{\rm rest}_{\chi_1\to \chi_2 \f} \frac{M_1}{T}.\eeq  
Only a fraction of 
\beq\laq{timeratio}
\frac{\D t_{\rm ignition}}{\D t_{\rm decay}}\sim \frac{g_\f \eta^3}{g_{\chi_1}}\frac{M_1^4}{4T^4} \(\ll 1\),
\eeq
 of $\chi_1$ decays. 
Although the slow decay with a small branching fraction of $p_\f^{\rm burst}/T$ decays into $\f$ with momentum $p_\f \sim p_\f^{\rm burst}$, it can fill the occupation number in the low momenta modes within a short period because of the small phase space volume $\sim g_\f\(p_\f^{\rm burst}\)^3$. 
At the end of this stage characterized by $f_{\f}(p_\f\sim p_\f^{\rm burst})\sim 1,$ or $t-t_i\sim \D t_{\rm ignition}$ we have a small occupation number for $p_{\f,\chi}\sim T$, $f_{\f}[p_\f \sim T], f_{\chi_2}[p_{\chi_2}\sim T]\ll 1,$ because of \eq{timeratio}.

The numerical result\footnote{Throughout the paper, the Boltzmann equation is solved on the lattices of the momenta, $\{\log{\hat{p}_\f}, \log{\hat{p}_{\chi_2}\}}$, in relevant ranges by using {\tt Mathematica}. } for this stage is shown in red shaded region in Fig.\ref{fig:para}, where I plot the solutions in the $[p_\f/T$ vs $f_\f], [p_\f/T$ vs $(p_\f/T)^3f_\f], [p_{\chi_2}/T$ vs $f_{\chi_2}]$ planes in the three panels from top to bottom, with taking $M_1/T=1/10, \eta=1,\D t_{\rm ignition} =\D t_{\rm decay}/2500 \ll \D t_{\rm decay}$, $t_i=0$, $\chi_1,\chi_2$ as Dirac fermions, and $\f$ as singlet scalar with $g_{\chi_{1,2}}=4, g_\f=1$. 
 In the top panel, the momentum modes around $p_\f/T\sim \O(0.001)\sim \eta M_1^2/(2T^2)$ grow to unity with $t\sim \D t_{\rm iginition}$. The timescale is much shorter than $\D t_{\rm decay}$. 

\subsection{Second stage: Burst} 
What happens afterward is a violent production of $\f$. 
This stage is characterized by \beq f_\f[p_\f \sim p_{\f}^{\rm burst}]\gtrsim 1, f_{\chi_2}(p_{\chi_2}\sim T)\ll 1,\eeq with which conditions, we have 
\beq 
\left.\frac{S}{f_{\chi_1}}\right|_{p_\f \sim p_{\f}^{\rm burst}} \sim f_{\f}[p_\f].  
\eeq
% is significant production of the $\f$ via the Bose enhancement. 
From \Eq{fphi}, we derive \beq \dot{f}_{\f}[p_\f \sim p_{\f}^{\rm burst}]\sim \frac{g_{\chi_1}}{g_\f} \frac{4T^3}{\eta^3M_1^3}  \Gamma^{\rm rest}_{\chi_1\to \chi_2 \f} f_{\chi_1} [p_{\chi_1}\sim T]f_\f[p_\f ].\eeq
By using the time-independent ~\eq{initial}, $f_\f[p_\f\sim p_{\f}^{\rm burst}]$ has exponential growth, thanks to the Bose enhancement.  
The growth rate is $\frac{g_{\chi_1}}{g_\f} \frac{4T^3}{\eta^3 M_1^3}  \Gamma^{\rm rest}_{\chi_1\to \chi_2 \f} f_{\chi_1} \sim\frac{ g_{\chi_1}}{g_\f} \frac{4T^3}{\eta^3M_1^3}  \Gamma^{\rm rest}_{\chi_1\to \chi_2 \f} \sim \D t_{\rm ignition}^{-1}$ where I used 
$f_{\chi_1}[p_{\chi_1}\sim T] \sim 1$.
Thus we get \beq \log{\(f_\f[p_\f \sim p_\f^{\rm burst}]\)} \sim \frac{t}{\D t_{\rm ignition}}. \eeq 
Therefore a burst production of $\f$ in the low momentum modes around $p_{\f}^{\rm burst}$ of $\f$ happens in a timescale not too different from $\D t_{\rm iginition}$. 

This stage can be found in the blue-shaded region in Fig.\ref{fig:para}, which is indeed characterized by the exponential growth of $f_\f(p_\f \sim M_1^2/T)$. Note that $t$ are changed with an interval $2\D t_{\rm ignition}$ for the plots here (rather than the exponential changes of the time for the plots in the previous stage). The growth rate is indeed $\sim \O(1)\times 1/\D t_{\rm ignition}.$
\subsection{Final stage: Saturation} 

The second stage is terminated due to the back reaction from the $\chi_2$ particles, which 
are simultaneously produced via the bose-enhanced $\f$ production. The relevant $\chi_2$ momentum in the reaction $\chi_1({ p_{\chi_1}\sim T})\to \f(p_\f \sim p_\f^{\rm burst}) \chi_2$ is $p_{\chi_2} \sim T$. Although the phase space volume of $\chi_2$ is much larger than that of $\phi$ modes around $p_\f\sim p_{\f}^{\rm burst}$, 
the exponential production of particles makes $f_{\chi_2}(p_{\chi_2}\sim T)$ soon reaches a quasi-equilibrium. 
The back reaction from $\chi_2$ stops a further burst production of $\f$. 
This equilibrium can be estimated by using $S\simeq 0$ with $f_{\f}(p_\f \sim p_\f^{\rm burst})\gg 1,$ which leads to 
%We get \WY{Check if this is correct for bosons}
\beq
\laq{eq}
f_{\chi_2}(p_{\chi_2}\sim T)\simeq  f_{\chi_1}(p_{\chi_1}\sim T)
\eeq
With \eq{initial}, the number density of $\chi_2$ at this stage is 
\beq
n_{\chi_2} \sim g_{\chi_2}\int_{p_{\chi_2} \sim T} \frac{d^3p_{\chi_2}}{2\pi^2} f_{\chi_2} \sim g_{\chi_2}\frac{T^3}{\pi^2}.
\eeq
From \Eq{conserv} and \Eq{tini} we arrive at 
\beq
\laq{num}
\boxed{n_{\f}[p_{\f}\sim p_\f^{\rm burst}]=n_\f^{\rm burst} \sim g_{\chi_2}\frac{T^3}{\pi^2}}.
\eeq
This form is similar to that from thermal distribution, $\sim g_\f T^3/\pi^2$, but it is different because the internal degrees of freedom, $g_{\chi_2}$, is  for $\chi_2$ and, importantly, it is composed of the low momentum modes, $p_\f \sim p_\f^{\rm burst} \ll T.$
I also showed that up to the saturation, the time is only passed by a few $\D t_{\rm ignition}$, 
therefore, we get the timescale for the burst production process to complete within 
\beq
\boxed{\D t_{\rm burst}\sim \O(1)\D t_{\rm ignition}}
\eeq
In the following analytical estimation, I neglect the short duration of the second stage, and I will use $\D t_{\rm ignition}$ to approximate the timescale to reach the final stage.

The stage discussed here is shown by the plots in the blue-shaded region. They overlap strongly because the system reaches a quasi-equilibrium. 
In addition, we numerically checked that $n_{\chi_2}[>0.5 T]\approx n_\f [<0.01 T] \approx 0.36 T^3$ at $t=15 \D t_{\rm ignition}.$ 
Here, $n_{i}[>p_{\rm cutoff}]\equiv g_i \int_{p_{\rm cutoff}}^{\infty}{\frac{d p_i p_i^2}{2\pi^2} f_i(p_i)},
n_{i}[<p_{\rm cutoff}]\equiv g_i \int_0^{p_{\rm cutoff}}{\frac{d p_i p_i^2}{2\pi^2} f_i(p_i)}.$ 
%Here we use $n_\f^{\rm burst}=n_\f [<0.01 T]= n_\f [<\O(10) p_{\f}^{\rm burst}]$
As a consequence, we have confirmed that the thermal reactions can produce $\f$ modes around $p_\f^{\rm burst}\ll T$ violently until the number density reaches $\sim g_{\chi_2} T^3/\pi^2$.

\begin{figure}[!t]
\begin{center}  
   \includegraphics[width=75mm]{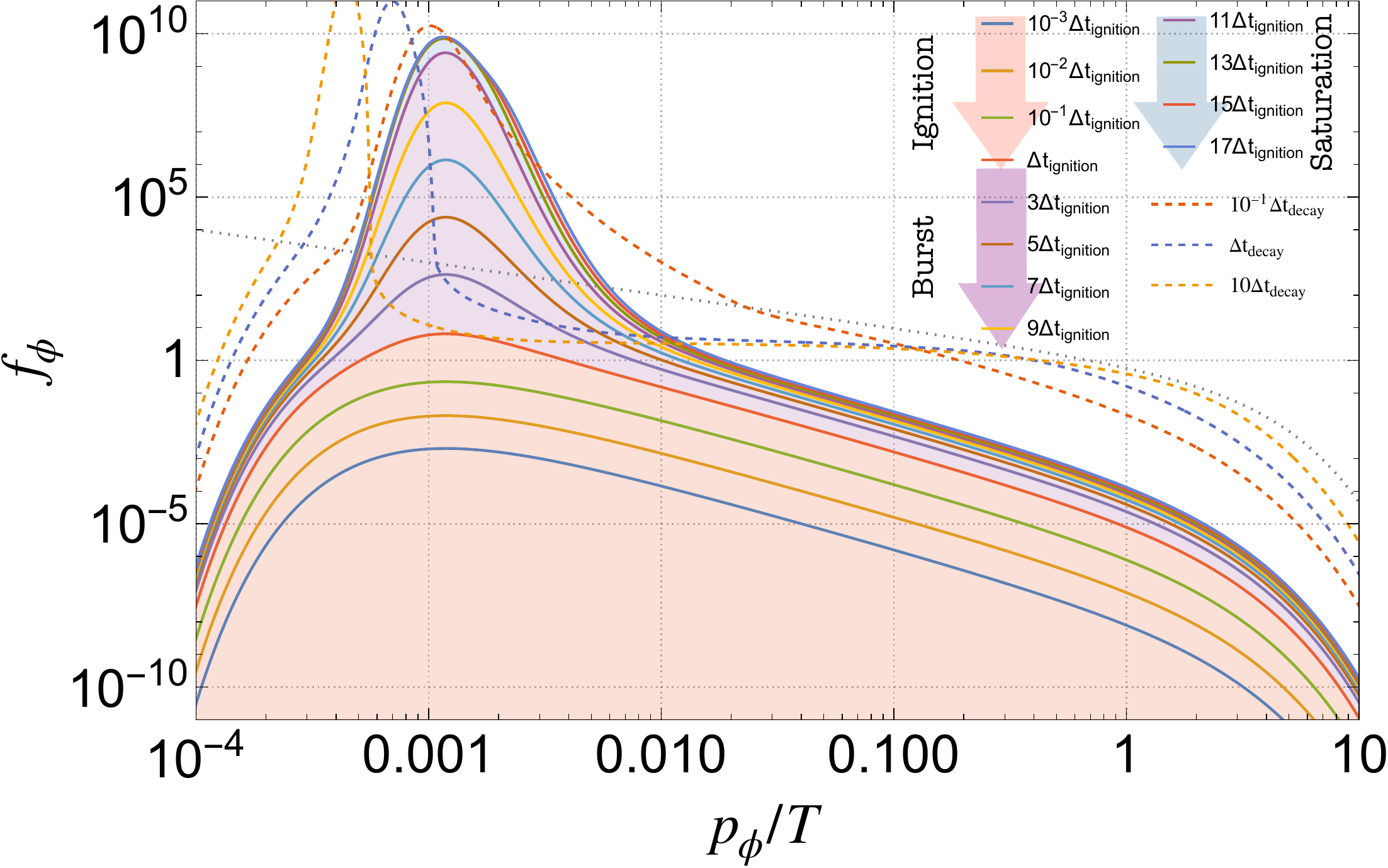}
      \includegraphics[width=80mm]{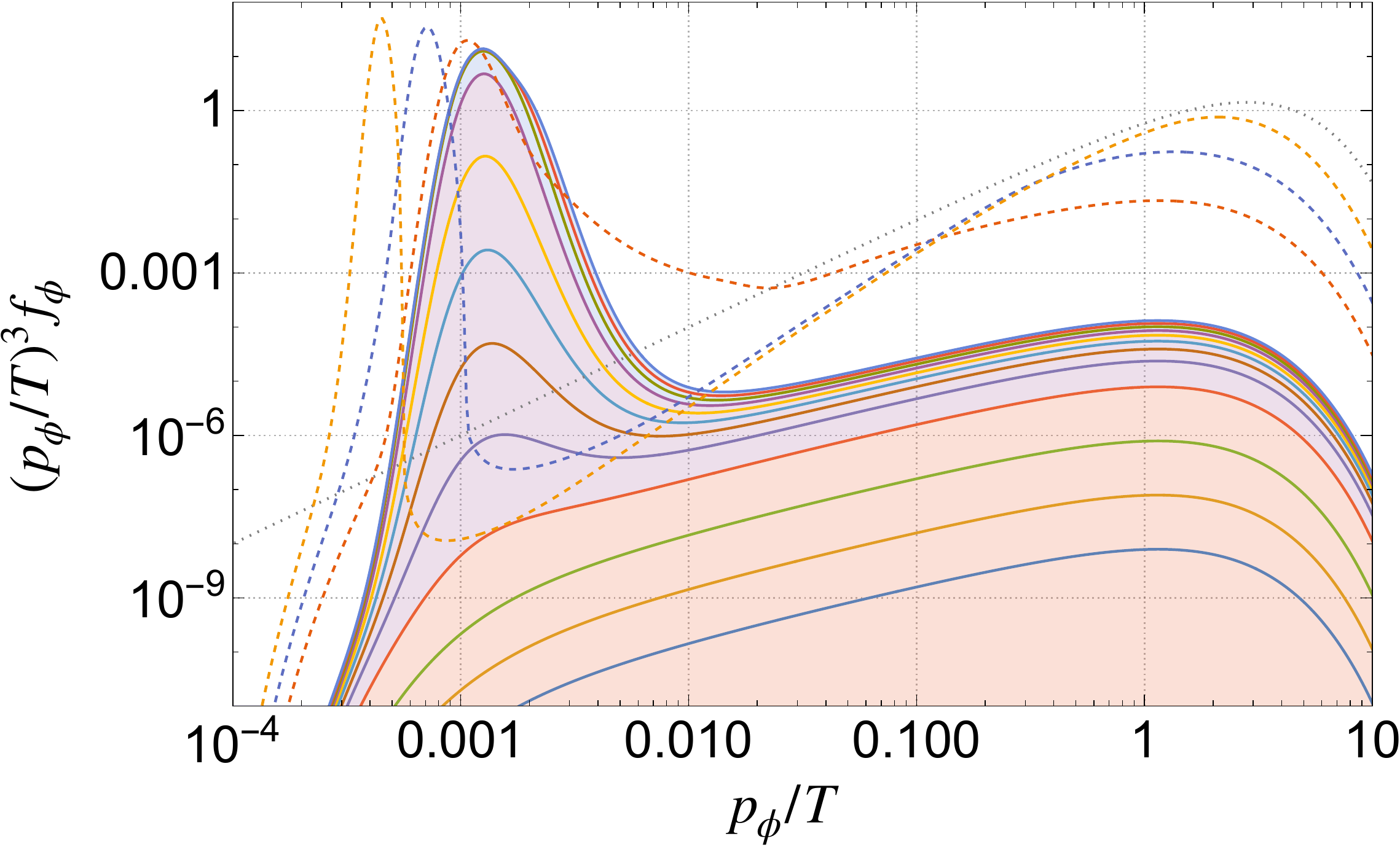}
            \includegraphics[width=80mm]{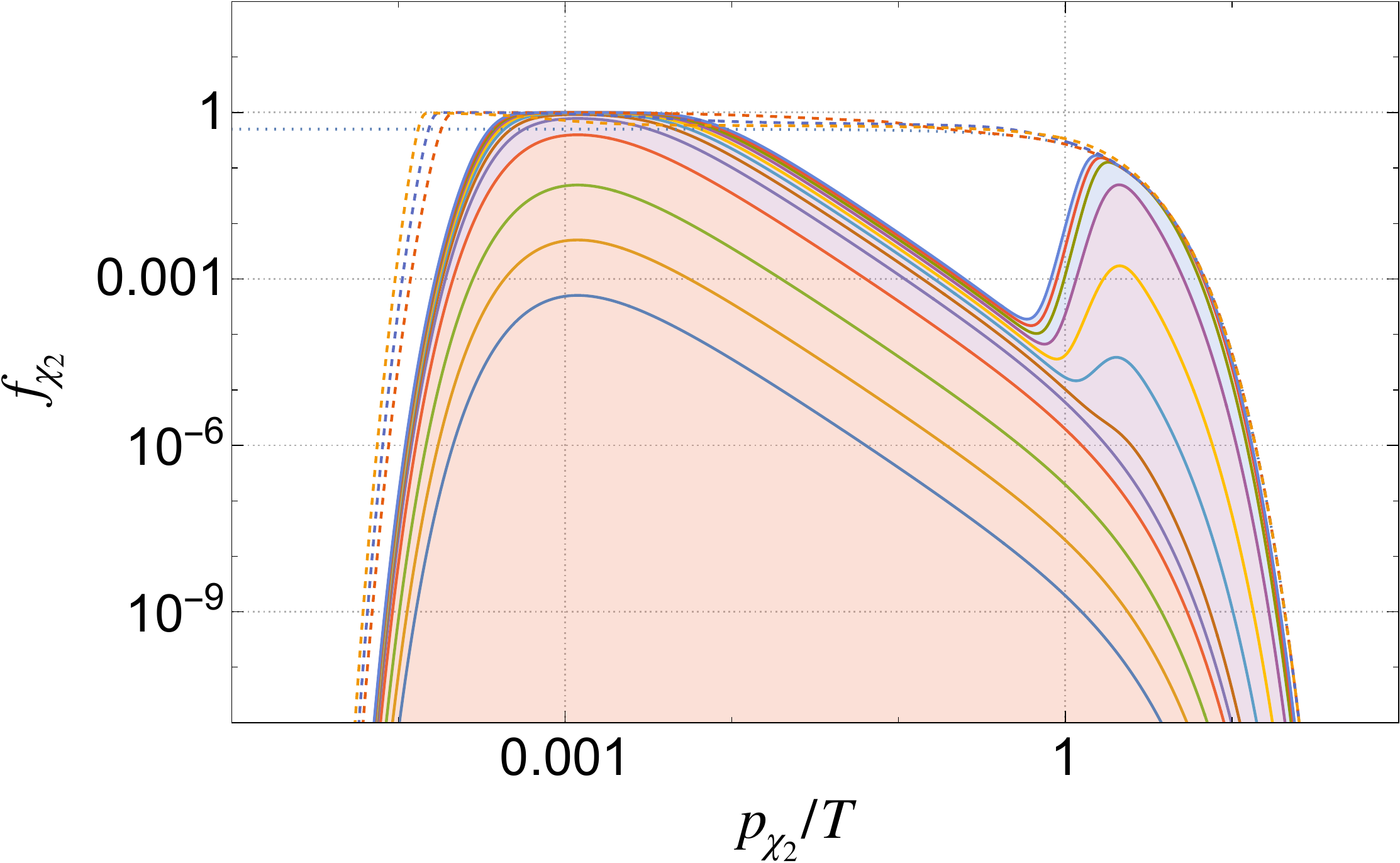}
\end{center}
\caption{The numerical solutions of the Boltzmann equation for $f_\f, f_{\chi_2}$ due to the process \Eq{decay} is shown in $f_\phi[t]-p_\f/T$ plane at several $t$ with $t_i=0$ [top panel]. The Hubble expansion is neglected  We take $M_1/T=1/10,M_2=0,$ and $\chi_{1,2}$ to be Dirac fermions with $g_{\chi_{1,2}}=4$, and $\f$ a scalar boson with $g_\f=1.$ The first stage, the ignition, of the burst production, is shaded in red with four plots for $t=\{10^{-3},10^{-2},10^{-1},1\}\D t_{\rm ignition}$ from bottom to top. The second stage, the burst, corresponds to the purple-shaded regime with four plots of $t=\{3, 5, \cdots 9\} \D t_{\rm ignition}$ from bottom to top. The final stage,  saturation, is found in the narrow blue shaded regime with four plots of $t=\{11, 13, \cdots 17\} \D t_{\rm ignition}$.  We also show for comparison that the plots with $t= \D t_{\rm decay}$ in dashed lines. 
Here $\D t_{\rm ignition}=2500\D t_{\rm decay}$ for the parameter set.  
In the middle panel, the solutions for $\(p_\f/T\)^3 f_\f$ is also shown. 
In the bottom panel, we display the solutions in $f_{\chi_2} -p_\f/T$ plane in the same setup and the time choices.   } 
\label{fig:para}
\end{figure}

\subsection{Slow thermalization after burst, and initial condition dependence }
\lac{thermalization}
In  Fig.\ref{fig:para}, I also displayed the distribution functions with $t\gg t_{\rm ignition}$ in the dashed lines. 
In the middle panel, the number density ($\propto$ the areas below the lines) around $p_\f^{\rm burst}$ are suppressed. 
Although in the next section, I will consider the parameter region that the physics at this timescale is seriously changed due to the expansion of the Universe, let us discuss the evolution in the flat Universe for the understanding of the stability of the quasi-equilibrium reached by the final stage of the burst production. 

What is happening is the usual thermalization via the decay/inverse decay at the timescale of $t\sim \D t_{\rm decay}$.
At this timescale, an $\O(1)$ fraction of plasma of $\chi_1$ of energy $\O(T)$ decays into $\chi_2$ and $\f$ with energies of $\O( T/2)$. Thus $f_{\chi_2}(p_{\chi_2}\sim T/2) \AND f_{\f}(p_\phi\sim T/2)$ tend to increase. This process did not reach an equilibrium so far because the burst production does not produce $f_\f(p_\f\sim T/2)$.
This happens much after the burst production. From the large hierarchy of the timescale \eq{timeratio} with $T\gg M_1$, %at every decay of $\chi_1$ into $T$ modes, 
the inverse decay via the burst process, $\f(p_\f \sim p_\f^{\rm burst}) \chi_2(p_{\chi_2}\sim T)\to \chi_1(p_{\chi_2}\sim T)$ happens immediately compensating the usual decay, i.e., it decreases $n_\f(p_\f\sim p_\f^{\rm burst})$ immediately to keep the quasi-equilibrium~\eq{eq}.

The phenomena discussed here can be seen from the dashed lines with $t=\{0.1, 1 ,10\}\D t_{\rm decay}$  in Fig.~\ref{fig:para}.  
Strictly speaking, the decrease happens from larger $p_{\f}$ which corresponds to larger $M_1^2/p_{\chi_{1},\chi_2}$, which corresponds to the faster boosted decay rate, $ \Gamma_{\chi_1\to \f \chi_2}M_1/{E_{\chi_1}}$. 
With the exponential hierarchy of timescale \eq{timeratio},  $f_\f(p_\f \sim  M_1^2/4p_0 \ll p^{\rm burst}_\f)$ at the moment $t\sim \D t_{\rm decay}$ has the production due to ignition with a timescale suppressed by the Boltzmann factor of $f_{\chi_1}( p_{0} \gg T) \sim \exp{(-p_{0}/T)}$ in the reaction $f_{\chi_1}(p_0 )\to f_{\chi_2}(p_{\chi_2}\sim p_0) f_{\f}(p_{\f}\sim  M_1^2/4p_0).$
This is the reason why the deeper IR modes are still populated at $t\sim  \D t_{\rm decay}$ in the top and middle panels of Fig.~\ref{fig:para}. Since the usual thermalization for the deeper UV modes of $\chi_1,\chi_2$, corresponding to the deeper IR modes of $\f$, is suppressed by the Lorentz factor, and Boltzmann suppression $\sim \exp{(-2p_{0}T)}$ for $f_{\chi_1}(p_{\chi_1}\sim  2p_{0})\to f_{\chi_2}( p_{\chi_2}\sim p_0), f_{\f}(p_\f\sim p_0)$ for the usual thermalization process, 
eliminating the deeper IR mode of $\f$ requires a longer timescale than $\D t_{\rm decay}.$\footnote{With this kind of suppressed IR modes, we can produce heavier DM than eV range, coldly, by explaining the measured DM abundance. \label{ft:2}}

The lesson we have learned here is that if the usual thermalization of $\chi_2$ at $p_{\chi_2}\sim T$ occurs, the burst-produced $\f$ in the IR modes is more-or-less eliminated to maintain the quasi-equilibrium \eq{eq}. This phenomenon may also happen for the thermalization of $\chi_2$ via the other interactions if they exist. \\

So far, we have discussed the case that $f_{\chi_2}=f_{\f}=0$ as the initial condition. Let me also comment on the numerical results for other initial conditions. 
I have checked that if we take $f_{\chi_2} \propto f_{\chi_2}^{\rm eq},\AND f_{\chi_2}\gtrsim f_{\chi_1}$ initially, the burst $\f$ production does not happen. 
On the contrary, it is also checked that even if $\f$ initially has a thermal distribution, we have the burst $\f$ production if $f_{\chi_2}$ is smaller than $f_{\chi_1}.$ They can be well understood by using the quasi-equilibrium~\eq{eq} and the number conserving feature \Eq{conserv} of the burst production via \eq{decay}.

\section{Cosmology of DM burst production}
Let us apply the burst production mechanism of the bosonic particles studied in the previous \Sec{3} to produce light DM because the burst-produced particles have $p_\f^{\rm burst}\ll T$. 
To discuss the burst production in a more realistic case, let me redefine
\beq
p_\f^{\rm burst}\equiv p_\f^{\rm burst}[T(t)]= \frac{\eta M_1^2}{2T(t)}
\eeq
here and hereafter. It has the same form as the previous section's $p_\f^{\rm burst}$, but I introduced the time dependence in $T[t]$ in $p_\f^{\rm burst},$ taking account of the Hubble expansion or thermalization of $\chi_1$.
In \Sec{4-1}, I remove the assumption of the flat Universe and assume the temperature $T[t]\propto a[t]^{-1}$ to show that the burst-produced bosons remain afterward. Then we estimate the DM abundance and discuss some model-independent constraints.  Since the production era of the DM is at the highest temperature of $T[t]$ in the regime $T[t]\propto a[t]^{-1}$, this production depends on the UV scenario of the radiation-dominated Universe. 
In Secs. \ref{chap:reheating} and \ref{chap:slowthe}, I will consider the scenarios that the DM produced during the reheating and through the thermalization of $\chi_1$, respectively. In Sec. \ref{chap:model} I will discuss the model-building for this production mechanism.

\lac{4}
\subsection{DM burst production in radiation-dominated Universe }
\lac{4-1}

To produce the DM, we need to guarantee that the number density of $\f$ via the burst production is not eliminated in the later history of the Universe. 
This is naturally achieved due to the expansion of the Universe, in which the momenta of free-particle redshifts, 
while the mass $M_i$ and $\G^{\rm rest}_{\chi_1\to \chi_2 \f}$ do not. 
In this section, let us further consider the case in the radiation-dominant Universe by assuming that the burst production timescale or $\D t_{\rm iginition}^{-1}$ is much faster than $H$, at \beq t=t_i=t_{\rm prod}\eeq which is our initial time for the discussion. 
The temperature is 
\beq
T[t_{\rm prod}]=T_{\rm prod}.
\eeq
I further consider that $\D t_{\rm decay}^{-1}$ is much smaller than the Hubble parameter at $t=t_{\rm prod}$. Then the burst production occurs because a Hubble time has many $\D t_{\rm ignition}$, and to discuss the burst, we can neglect the Hubble expansion resulting in the essentially same setup as discussed in the previous section. 
Afterwards the thermal distribution of $\chi_1$ has a time-dependent temperature scaling as 
\beq T[t]= T_{\rm prod} \frac{a[t_{\rm prod}]}{a[t]}.\eeq 
I assumed there is no entropy production or dilution to increase or decrease $T a$. 
Thus the typical momentum for the burst production blueshifts, i.e., $p_\f^{\rm burst}\propto a,$ but particle momentum redshifts, $p_\f\propto a^{-1}$. Within one Hubble time, the momentum of produced light DM before at $t_{\rm prod}$ soon becomes smaller than $p_\f^{\rm burst}[t]$, 
i.e., 
\beq
p_\f^{\rm burst}(t_{\rm prod}) \frac{a[t_{\rm prod}]}{a[t]}< p_\f^{\rm burst}(t) =p_\f^{\rm burst}(t_{\rm prod}) \frac{a[t]}{a[t_{\rm prod}]}.
\eeq
 with $a[t]>a[t_{\rm prod}]$ due to the redshift and blueshift. 
  Since the production/destruction rate of the modes with $p_\f \ll p_\f^{\rm burst}$ via \Eq{decay} is Boltzmann suppressed by $f_{\chi_1}\sim e^{-\eta M_1^2/(2p_\f T)}$ (see \Eq{fphi}), the DM production/destruction for the mode produced at $t=t_{\rm prod}$ will be kept intact later thanks to the expansion of Universe. 

The production of modes around $p_\f \sim p_\f^{\rm burst}[t]$ later is also suppressed since $f_{\chi_2}(p_{\chi_2}\sim T)$ reaches the quasi-equilibrium $f_{\chi_1}(p_{\chi_1}\sim T)\sim f_{\chi_2}(p_{\chi_2}\sim T)$ at the first short moment $t\approx t_{\rm prod}$. This is because  $p_{\chi_1}, p_{\chi_2}\propto a^{-1}, T\propto a^{-1}$ later. In other words, once $\chi_2$ has the number density $\sim g_{\chi_2}T^3/\pi^2$, which is close to the upper bound from the thermal production, \Eq{conserv} also sets the upper bound of the number density of $\f$. Therefore once $n_\f\sim n_\f^{\rm burst} \sim g_{\chi_2} T_{\rm prod}^3/\pi^2$ is fulfilled due to the burst production, the reaction is afterward suppressed. 
%I emphasize that even if at a later time $\D t^{-1}_{\rm decay}\gg H$, the burst-produced modes are not eliminated because of the kinematics (and Boltzmann suppression). 
A numerical simulation for a similar setup is shown in Fig.~\ref{fig:2} by assuming a phase of reheating followed by  the radiation-dominated Universe (see for a detailed explanation of the figure in \Sec{reheating}). We see the burst-produced component indeed is frozen at a much later time at which the $p_\f\sim T$ modes are mostly thermalized. This is a very different point from the case in flat Universe \Sec{thermalization}.

To sum up, the condition for the burst production to occur in the radiation dominated Universe is 
 \beq\laq{cond} \D t_{\rm ignition}^{-1}\gg H \gg \D t_{\rm decay}^{-1} {\rm ~at~}T=T_{\rm prod} \eeq 
The first inequality is that the Hubble expansion can be neglected compared with the timescale for the burst production.
The second inequality is for suppressing the usual thermalization eliminating the burst-produced $\f$. This is because with $\D t_{\rm ignition}^{-1} ,\D t_{\rm decay}^{-1}\gg H,$ the setup by neglecting the Hubble expansion will be essentially the same as in \Sec{thermalization} (with additional interaction for $\f,\chi_2$, the additionally induced thermalization of $\chi_2$ and $\f$ should probably be also smaller than the Hubble rate [see {\Sec{slowthe}]). If this is satisfied the DM is produced with \eq{num} with $T=T_{\rm prod}.$ 

One notices with the assumption $T\propto a^{-1}$ and $H\propto a^{-2}$, the condition~\eq{cond} is more likely to satisfy in an early time. %Thus we may have other component for the $T=T_{\rm prod}$ should be the highest temperature that the setup is justified. 
In other words, the production should be UV scenario dependent. Since in the following subsections, we will focus on some natural scenarios that the discussion here is applicable by properly choosing $T_{\rm prod}$, let us continue our discussion. 

\paragraph{DM abundance and the mass range}

Once the condition \eq{cond} is satisfied, later, the ratio of the burst-produced DM number density to plasma entropy density conserves until today. 
 The cold component of the abundance can be estimated from 
\beq \Omega_\f  = m_\f \left. \frac{n^{\rm burst}_\f}{s}\right|_{T=T_{\rm prod}}\frac{s_0}{\rho_c}.\eeq 
Here $\rho_c,s_0$ are the present critical density and the entropy density, respectively, 
and $n^{\rm burst}_\f \sim g_{\chi_2} T_{\rm prod}^3/\pi^2$ (see \Eq{num}),$s=g_{\star,s}\frac{2\pi^2}{45} T_{\rm prod}^3.$ 
$g_{\star,s}$ is the relativistic degrees of freedom for entropy, ($g_{\star}$ will be used as the degrees for energy density). $g_{\star,s}$ should include $(7/8)^f (g_{\chi_1}+g_{\chi_2})$ with $f=1(0)$ for fermionic (bosonic) $\chi_{1,2}$, because they are relativistic soon after the burst. 
In the following, I assume that the comoving entropy carried by $\chi_{1,2}$ is released to the lighter SM particles  before the neutrino decoupling. In addition, $\chi_{1,2}$ are supposed not to dominate the Universe during the thermal history (no further entropy production other than $\sim g_{\chi_1} T^3, g_{\chi_2} T^3$).   %\footnote{There is also a possibility that the comoving entropy stored in $\chi_{1,2}$ is released (remain) in a dark sector contributing to dark radiation.}
By requiring the $\f$ abundance $\Omega_\f   $ equal to the measured DM density~\cite{Planck:2018vyg}
$\Omega_{\rm DM}\approx 0.26,$
I, therefore, get 
\beq
\laq{abundance}
m_\f = 50 {\rm eV} \frac{4}{g_{\chi_2}}\frac{g_{\star,s[T_{\rm prod}]}}{100}.
\eeq
Since $g_{\star,s}$ include $g_{\chi_2},$ at the $g_{\chi_2} \to \infty$ limit, this reduces to the lower bound of the mass of $\f$, 
while we may also have an upper bound by restricting $g_{\star,s}\lesssim \O(100)$:
\beq\laq{massrange}
2\EV \lesssim m_\f\lesssim \O(100)\EV
\eeq
is the generic prediction.\footnote{If there is a large amount of entropy production after the production by, e.g., $\chi_1,\chi_2$ dominating the Universe, the DM can be heavier, which is not taken into account here.}

\paragraph{Constraints from structure formation}
To have successful structure formation, we need the free-streaming length of the DM to be sufficiently suppressed. The 
free-streaming length of the cold component can be estimated by using, 
\beq
L_{\rm FS}=a_0 \int^{t_{\rm eq}}{dt{v_{\f}[t]}}.
\eeq 
Here $v_{\f}$ is the typical physical velocity of $\f$ DM, $a_0$ is the present scale factor, $t_{\rm eq}$ is the time at matter-radiation equality. 
By approximating $v_{\f}\sim \frac{p_{\f}^{\rm burst}(t_{\rm prod}) a[t_{\rm prod}]/a}{\sqrt{\(p_{\f}^{\rm burst}(t_{\rm prod}) a[t_{\rm prod}]/a\)^2+m^2_\f}}$, the bound $L_{\rm FS}<0.06 $Mpc \cite{Viel:2005qj, Irsic:2017ixq}
 (see a similar mapping for heavy DM from inflaton decay \cite{Li:2021fao})
leads to
 \beq
 \laq{Lya}
\sqrt\eta\frac{M_1}{T_{\rm prod}}\lesssim 0.02 \(\frac{g_{s\star}[T_{\rm prod}]}{100}\)^{1/6} \sqrt{\frac{m_\f}{\EV}}.
 \eeq
The required hierarchy is not too large.\footnote{{Therefore, I will not discuss the model-dependent issues, e.g., the coherent scattering and the perturbatively of the Boltzmann equation, that are important when the occupation number is very large.   }}

\paragraph{Suppression of the hot components}
Strictly speaking, other than the burst-produced component, production of $\f$ may occur via the usual decay/inverse decay process. 
This implies we may have a mixed DM after the decoupling of $\f$ from the thermal plasma
\beq
n_\f^{\rm tot}[t]=n_\f^{\rm burst}[t]+ n_\f^{\rm th}[t]
\eeq
where the first term denotes the part from the burst produced component, which is dominated by the momentum mode $p_\f^{\rm burst}\sim \eta M_1^2 a[t_{\rm prod}]/(2T_{\rm prod} a[t]) \ll T[t]$, while the latter one is from the 
ordinary thermal production via \Eq{decay}. We have neglected the latter component so far in discussing the free-streaming length.  
%In the eV range, the hot component contributes to a fraction of the hot DM. 
Indeed, the hot component $n_\f^{\rm th}$ of $\f$ should be suppressed to be below $\O(1-10)\%$ level depending on the mass range~\cite{Diamanti:2017xfo}. 
Indeed in the numerical simulation for Fig.~\ref{fig:2}, I get $n_{\f}^{\rm th}/n_{\f}^{\rm tot}\equiv \frac{n_\f [<0.1 T]}{n_\f[<0.1T]+n_\f [>0.1 T]}|_{t=5\times 10^5 t_R}\sim 30 \%$ for $\chi_{1,2}$ being the Dirac fermion case (top panel), and dominant hot components for $\chi_{1,2}$ being singlet scalars (middle panel). 
They may be in tension with the constraint. 
Here let me point out three kinds of parameter regions with suppressed hot DM components.

First, we can suppress the hot DM component by considering $\eta \lesssim 1$, i.e., the mass of $\chi_2$ is non-negligible. The contribution can be analytically estimated by using $n_{\f}^{\rm th}\sim \eta^3 T^3/\pi^2$, because the momentum of $\f$ can be at most produced up to $\eta T$~(see \Eq{mom}). We expect a near thermal distribution at  $p_\f \lesssim \eta T$ while the spectrum is suppressed at $p_\f \gtrsim \eta T$ compared to the thermal one. For instance, with $\eta=1/2$, in the bottom panel of Fig.~\ref{fig:2}, the thermal component is indeed suppressed. 
%We expect what is happening around $t=5\times 10^{3}t_R$ is the inverse decay of $f_\f(p_\f\sim T) f_{\chi_2}(p_\chi \sim 1/\eta T)\to f_{\chi_2}(p_{\chi_2} \sim 1/\eta T),$ which is boltzman suppressed.  to 
We checked $n_{\f}^{\rm th}/n^{\rm tot}_{\f}\sim 3\%$. In this case, $p_{\f}^{\rm burst}$ is also suppressed~(see the Figure c.f. \Eq{pburst}). 

The second possibility is to consider the small $m_\f$ range by taking $g_{\chi_1}/g_{\star,s}$ large in \Eq{abundance}. The effects have already been seen by comparing the top and middle panels in Fig.~\ref{fig:2}. 
If the mass is below $5-10\EV$ with $T_{\rm prod}\gtrsim 100\GEV$, we can have a successful cold DM scenario together with the hot DM similar to the scenarios~\cite{Daido:2017wwb, Daido:2017tbr}. 
This scenario may be naturally justified if $\chi_i$ are charged under some non-abelian group.\footnote{\label{ft:1}In the scenarios solving the quality problem by using a non-abelian gauge group, multiple PQ Higgs bosons/singlet fermions with similar masses may be predicted~\cite{DiLuzio:2017tjx, Lee:2018yak, Ardu:2020qmo, Yin:2020dfn}. Also, the scenario generically predicts fermions/bosons with large internal degrees. There may also be additional neutral bosons in the scenarios enhancing the sphaleron rate for baryogenesis with lower reheating temperature than electroweak scale~\cite{Jaeckel:2022osh} and the model having QCD axion DM around eV with larger decay constant than the conventional one~\cite{Agrawal:2017ksf}. }

The last possibility is that $\D t^{-1}_{\rm decay}$ never becomes faster than $H$ until $T\sim M_1.$ 
This may be considered as the ``freeze-in" scenario in which the usual interaction rate with the plasma for $\f$ is always smaller than the Hubble expansion. Then the usual thermal production of $f_{\f}(p_\f\sim T)$ is always suppressed, and thus $n^{\rm th}_\f/n_\f^{\rm tot}$ is suppressed. One numerical example is shown in Fig.\ref{fig:3}, where the hot component is suppressed to be ${n_\f^{\rm th}}/{n_\f^{\rm tot}}\sim 5\%$ (See \Sec{reheating} for detail of the figure).

I also comment that the hot DM bound disfavors the simple scenario $\chi_2=\f$. 
This is because $f_{\f}[p_\f\sim T]=f_{\chi_2}[p_\f\sim T]$ is also obtained via the burst production (as seen from the middle panel of Fig.\ref{fig:2}). Any of the above possibilities may not be useful for this case.

 \begin{figure}[!t]
\begin{center}  
   \includegraphics[width=75mm]{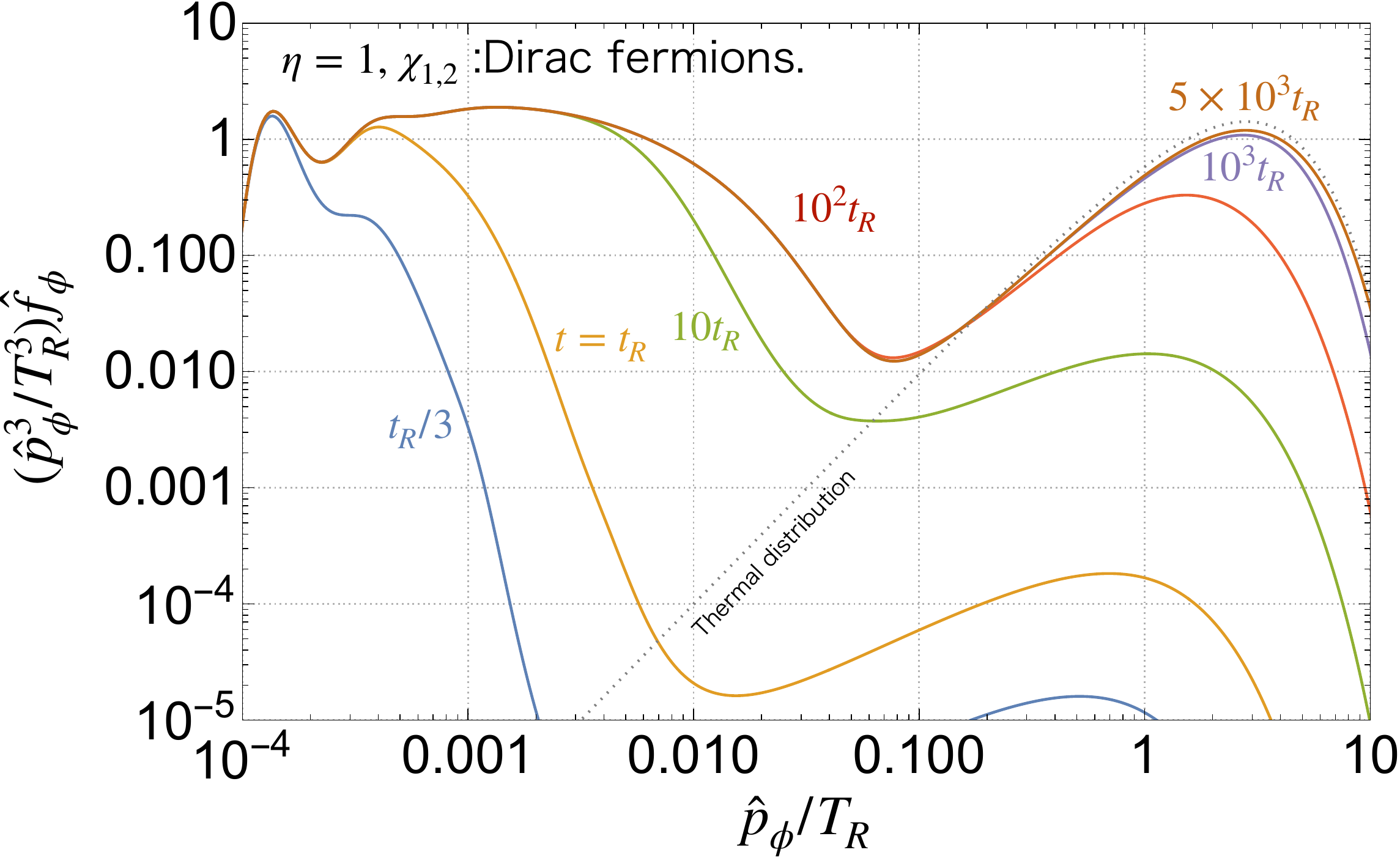}
      \includegraphics[width=75mm]{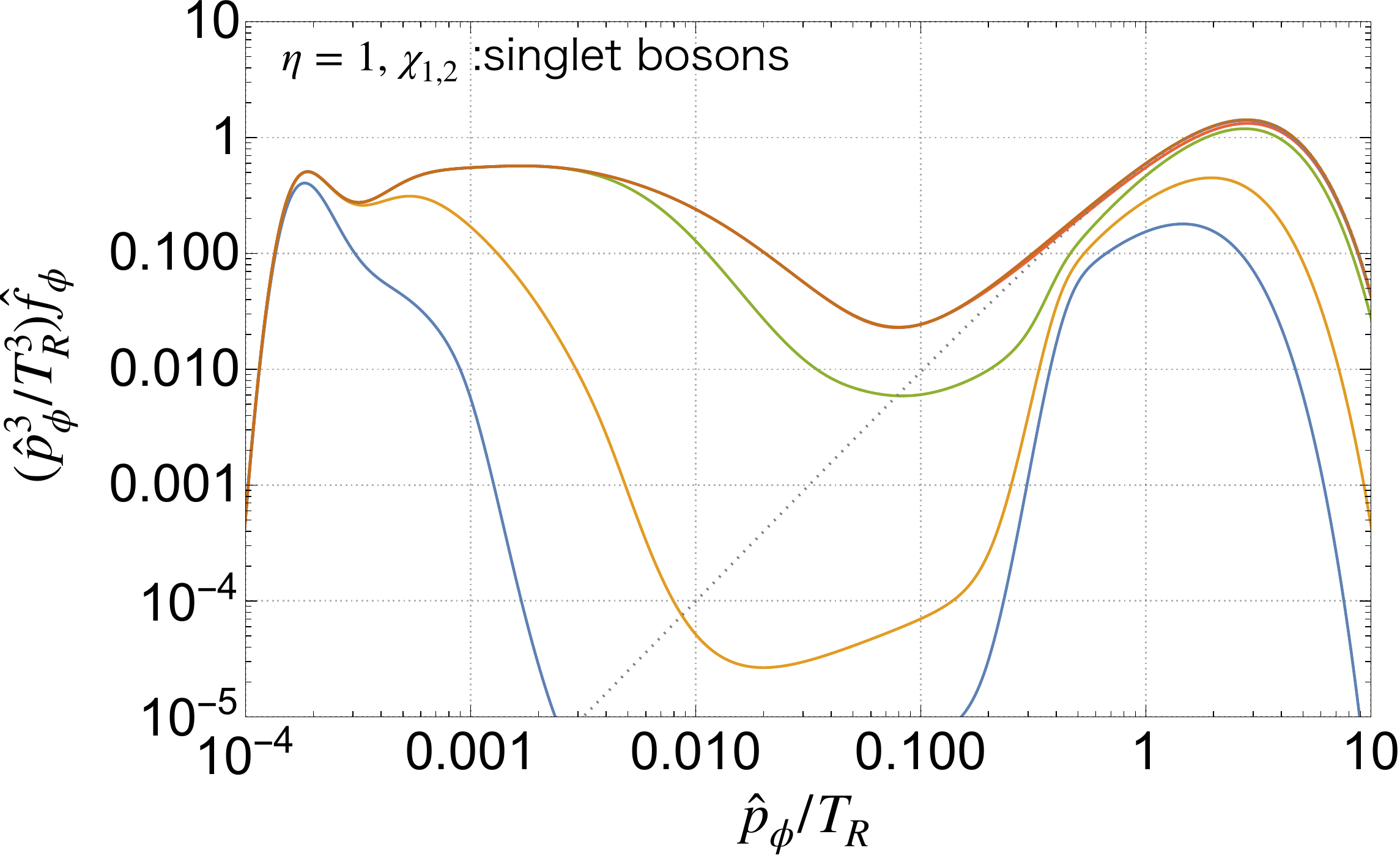}
            \includegraphics[width=75mm]{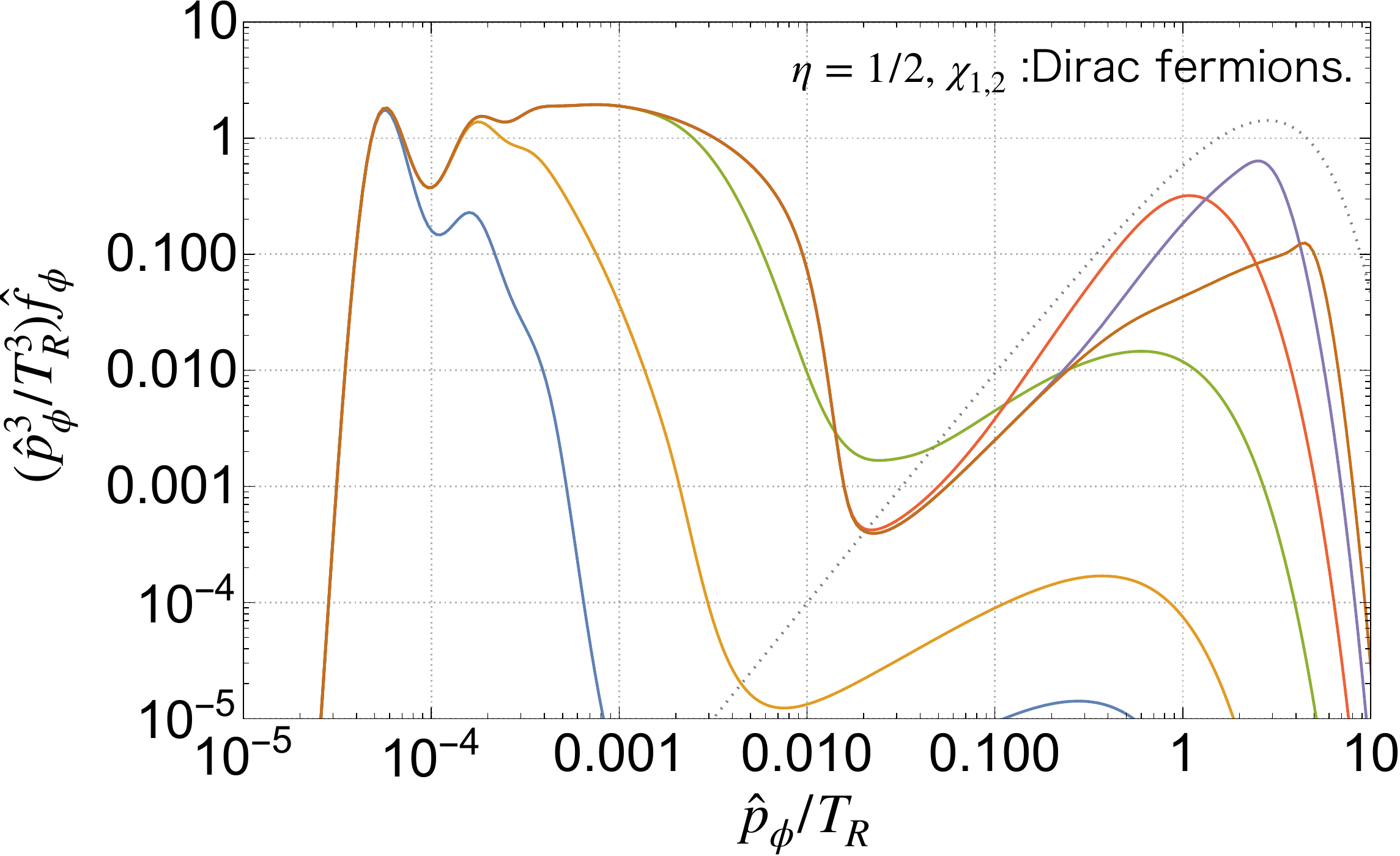}
\end{center}
\caption{The numerical simulation of DM distribution function $\hat{p}_\f^3T_R^{-3} \hat{f}_\phi$ in expanding Universe by varying $\hat{p}_\f/T_R $ with $t=\{1/3,1,10,10^2,10^3, 5\times 10^3 \}t_R$. The initial cosmic time is $t_i=1/10 t_R$ at which $\hat{f}_{\chi_2}=\hat{f}_\f=0$ is set. $\Gamma_{\chi_1\to \chi_2\f}=10^{-3} t^{-1}_R, M_1=T_R/10 .$ We take $g_{\chi_{1,2}}=4$ with $\chi_{1,2}$ being fermions in the top panel. 
The dotted line shows the thermal distribution for $\f$. $\hat{p}_i \equiv p_i a[t]/a_R.$ In the middle and bottom panels, cases with $\eta=1,\AND 1/2$ with $g_{\chi_{1,2}}=1 \AND 4$ and $\chi_{1,2}$ being singlet scalars and fermions are shown, respectively. Other parameters/variables are the same as in the top panel. In all panels, I consider $\f$ as a scalar with $g_{\f}=1.$
}
\label{fig:2}
\end{figure}

 \begin{figure}[!t]
\begin{center}  
   \includegraphics[width=75mm]{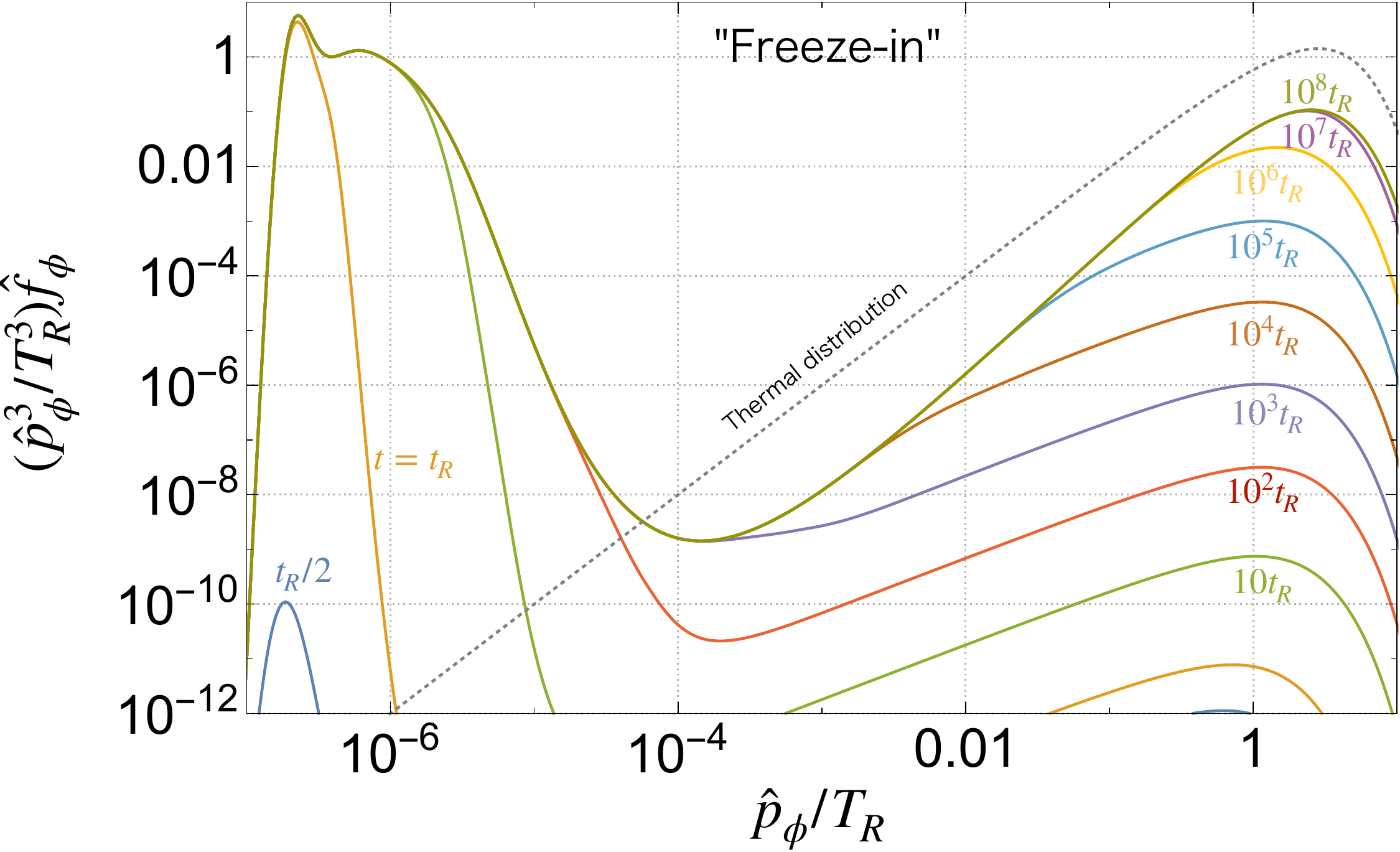}
\end{center}
\caption{The numerical simulation for the ``freeze-in" scenario that the usual thermalization rate of $\f$ is always slower than the Hubble rate. 
Setups are the same as the top panel in Fig.\ref{fig:2}. I take $t_i=1/3 t_R$, $M_1=T$ at $t=2.5\times 10^{5}t_R,$ at which $\Gamma_{\chi_1\to \chi_2 \f}^{\rm rest}=0.01 H.$ 
The plots are for $t=\{1/2,1,10,10^2,10^3, 10^4,10^5,10^6,10^7,10^8 \}t_R$ from bottom to top. }
\label{fig:3}
\end{figure}

\subsection{DM burst production during reheating} 
\lac{reheating}

One realistic possibility that the setup for the numerical simulation can apply is the DM production at the end of reheating. 
Given that $\chi_1$ is always thermalized, the burst production was found to be UV-dependent in the radiation-dominated Universe, which motivates me also to study the behavior during the reheating phase.
To be more concrete, let us focus on the case $\D t^{-1}_{\rm ignition}$ is faster than the Hubble expansion rate at the end of reheating $t=t_R$,
\beq
\laq{condreheating}
\D t_{\rm ignition}^{-1}|_{t=t_R}\gg H(t=t_R).
\eeq
$t=t_R$ is defined by $H=\sqrt{\rho_{\rm tot}/3M_{\rm pl}^2}=\G_{\rm reh}$ where $\Gamma_{\rm reh}$ is the decay rate of inflaton, moduli or other particle that is responsible for reheating, $\rho_{\rm tot}$ the total energy density of the Universe, $M_{\rm pl}\approx 2.4\times 10^{18}\GEV$ the reduced Planck scale. The cosmic temperature at this moment is defined as $T=T_R.$

 During the reheating, the radiation is contiuously produced via $\rho_r \sim \rho_{\rm tot} \Gamma_{\rm reh}/H$.  As conventionally, I assume the matter-dominated Universe during the reheating, $ H\propto a^{-3/2}, \rho_{\rm tot}\propto a^{-3}$, which gradually decays into radiation. Thus $T\propto \rho^{1/4}_r \propto a^{-3/8}$, i.e., the temperature of the plasma due to the reheating decreases slower than $a^{-1}$.  
 The ignition rate scales as
 \beq
\D t_{\rm ignition}^{-1}|_{t<t_R}\propto T^3\propto a^{-9/8}
\eeq
which decreases slower than the Hubble rate, i.e., during the reheating, the ignition rate is IR dominant.  
Given \Eq{condreheating}, the burst production rate is still larger than the Hubble rate if we go back in time for a while, during which we have the burst production. Indeed, to satisfy the quasi-equilibrium condition \Eq{eq} during the reheating phase, $\chi_2$ is gradually produced through \Eq{decay}, and $f_{\chi_2}(p_{\chi_2})$ also scales with $T\propto a^{-3/8}$. 
In the period with the quasi-equilibrium, $\chi_2$ has the comoving number density, $n^{\text{ quasi-eq.}}_{\chi_2}a^3 \propto a^{15/8}$. 
Thus I get
\beq
\dt \(n^{\text{quasi-eq}}_{\chi_2}a^3\)\propto H a^{15/8} \propto a^{3/8}.
\eeq
This increases in time during the reheating, the largest comoving number density is produced at the last Hubble time in the reheating phase. 
After the end of the reheating, the momenta on both sides of \Eq{eq} scales as $a^{-1}$, and the production of the comoving number density of $\chi_2$ is suppressed, as discussed in \Sec{4-1}.  
From \Eq{conserv}, the $\f$ IR modes are also produced gradually and most efficiently at the end of the reheating. 
Thus, we can choose 
\beq
\laq{Tprodreheating}
T_{\rm prod}\sim T_R 
\eeq
as a not-too-bad approximation.
 The typical momentum of $\f$ due to the burst production is then around $\frac{\eta M_1^2}{2T_R} \frac{a[t_R]}{a[t]}.$ 

 I simulate this scenario to get the $\f$ spectrum in Fig.~\ref{fig:2}. I used $a= a_R({t}/{t_0})^{1/2}\tanh{((t/t_0)^{1/6})}$ and $T= T_0 ({a_R}/{a}) \tanh{(a/a_R)^{5/8}}$, i.e. the reheating ends at $a=a_R\sim a[t_R], t=t_R$, in \Eqs{fphi}, \eq{fpsi}, and \eq{initial}. 
I displayed $\frac{p_\f^3{a^3[t]}}{T_R^3 a^3_R}f_\phi$ in expanding Universe by varying $p_\f a[t]/a_R T_R $ at $t=\{1/3,1,10,10^2,10^3, 5\times 10^3 \}t_R$ from bottom to top. The initial time is chosen as $t_i=1/10 t_R$ at which $\hat{f}_{\chi_2}=\hat{f}_\f=0$ is set. $\Gamma_{\chi_1\to \chi_2\f}=10^{-3} t^{-1}_R, M_1=T_R/10.$ I take $\eta=1$, $\chi_{1,2}$ are Dirac fermions (singlet bosons) with $g_{\chi_{1,2}}=4 (1)$ in the top (middle) panel, and take $\eta=1/2$ for $\chi_{1,2}$ being Dirac fermions with $g_{\chi_{1,2}}=4$ in the bottom panel. 
In Fig. \ref{fig:3}, the same setup as the top panel of Fig.\ref{fig:2} is plotted, but I take $t_i=1/3 t_R$,\footnote{To reduce the calculation cost, I took a relatively close initial time to $t_R$ for the end of reheating. There is a sharp peak, which should be understood as the burst produced $\f$ at $t\sim t_i$ with the initial condition $f_{\f,\chi_2}=0$, which seems to be dominant for the number density. This number is expected to be diluted if reheating lasts long, and the spectrum becomes UV insensitive (while the total amount of $n^{\rm burst}_\f a^3|_{t\gg t_R}$ does not change much if $n_{\chi_2}\sim g_{\chi_2}T^3/\pi^2$ after the reheating before the thermalization due to \Eq{conserv}). 
On the other hand, this $t_i \sim t_R$ simulation will be a not-too-bad assumption by considering some scenarios with a short reheating phase. For instance, the ALP inflation scenarios \cite{Takahashi:2019qmh, Takahashi:2020uio,Takahashi:2021tff} and some scenarios for baryogenesis after the supercooling~\cite{Azatov:2021irb,Baldes:2021vyz, Azatov:2022tii} (see also \cite{Azatov:2020ufh,Azatov:2021ifm}) have this feature. 
 } $M_1=T$ at $t=2.5\times 10^{5}t_R,$ at which $\Gamma_{\chi_1\to \chi_2 \f}^{\rm rest}=0.01 H,$ i.e. the $\D t_{\rm }^{\rm decay}$ is $\O(100)$ times longer than the age of the Universe when $\chi_1$ becomes non-relativistic. 
The plots are for $t=\{1/2,1,10,10^2,10^3, 10^4,10^5,10^6,10^7,10^8 \}t_R$ from bottom to top.

We can see in any case that $\f$ is burst-produced at the momentum mode $p_\f^{\rm burst}(T_R)/T_R= \eta M_1^2/(2T^2_R)\sim \eta \times \O(0.001),$ at the time $t=\O(1)t_R.$ Later, the cold component around the comving momentum $p_\f^{\rm burst}(t_R)a[t_R]/a[t]T_R$ is frozen later for exponentially long time.

From the above discussion, it is clear that the DM mass \eq{abundance} and coldness~\eq{Lya} can be estimated by using \Eq{Tprodreheating}.

\subsection{DM burst production during thermalization}
\lac{slowthe}
One can also consider the DM production during the thermalization of $\chi_1$. 
 Here, for generality, I use 
$\G_{\rm th,1},\G_{\rm th,2}\AND \G_{\rm th,\f}$, respectively, for denoting the thermalization rate for $\chi_1,\chi_2 \AND \f$ from some additional reactions than \Eq{decay}, like the scatterings with other SM particle plasma with temperature $T\propto a^{-1}$.  
I will neglect $\G_{\rm th,2} \AND \G_{\rm th, \f}$ in the main discussion, and I will come back at the end of this subsection. 
Initially I take $f_{\chi_1,\chi_2,\f}=0$, and $\G_{\rm th,1}<H$ at $t=t_i$. Furthermore, I assume the radiation-dominated Universe, and $\Gamma_{\rm th,1}/H \propto a^{q_{\rm th,1}}$  with $q_{\rm th,1}$ being a positive number. 
For instance, the thermalization via the renormalizable interactions with relativistic particles may have $\G_{\rm th,1}\propto T$ from dimensional arguments, leading to $q_{\rm th,1}=1$.
With those assumptions, we have a cosmic time $t_{\rm th,1}$ and cosmic temperature $T_{\rm th,1}$ that $\Gamma_{\rm th,1}=H$.

Although a more detailed study of thermalization relies on the momentum-dependent and model-dependent interactions of $\chi_1$, 
 let us assume that thermalization mainly occurs for the number density with the typical momentum $p_{\chi_1}\sim T$ for simplicity. 
 If $\chi_{1,2}$ are fermions, this should be a good approximation because the reaction that produces IR modes with small phase space volume is soon Pauli-blocked.  
 Thus $f_{\chi_1}(p_{\chi_1}\sim T)$ increases with time with \beq f_{\chi_1}(p_{\chi_1}\sim T)\sim \frac{\G_{\rm th,1}}{H}  {\rm~with~} t\ll t_{\rm th}.\eeq 
To estimate the ignition rate, let us take account of $f_{\chi_1}$ in the integral of \Eq{fphi}, and use
\beq
\D t_{\rm ignition}^{-1} \equiv \frac{g_{\chi_1}}{g_\f}\frac{T^3}{4\eta^3 M^3_1} \G^{\rm rest}_{ \chi_1\to \chi_2\f}\frac{f_{\chi_1}(p_{\chi_1}\sim T)}{f^{\rm eq}_{\chi_1}(p_{\chi_1}\sim T)}
\eeq
where $\frac{f_{\chi_1}(p_{\chi_1}\sim T)}{f^{\rm eq}_{\chi_1}(p_{\chi_1}\sim T)}$ was $1$ in the previous analysis since we assumed $\chi_1$ in the thermal equilibrium \Eq{initial}. 
In particular, we focus on 
the case that burst production rate is faster than the Hubble rate at the thermalization,\footnote{If this is not satisfied, but satisfied in the early stage of thermalization, $t\ll t_{\rm th,1}$ when $f_{\chi_1}<f_{\chi_1}^{\rm eq}$, we will have the suppressed $n_\f^{\rm burst}<g_{\rm \chi_2}T_{\rm prod}^3/\pi^2$ with $T_{\rm prod}$ being the cosmic temperature at $H= \D t^{-1}_{\rm ignition}$. Thus the DM mass to explain the DM abundance is enhanced. 
This is the case $q_{\rm th,1} < 1,$ because $t_{\rm ignition}^{-1} \propto a^{-3+q_{\rm th,1}}$ decreases faster than $H\propto a^{-2}$ does. \label{ft:3}
}
\beq 
\laq{condth}
\D t_{\rm ignition}^{-1} \gg H~ {\rm at}~ t=  t_{\rm th,1} .
\eeq
It scales as for $t\ll t_{\rm th,1}$
\beq
\D t_{\rm iginition}^{-1}\propto a^{-3+q_{\rm th,1}}.
\eeq 
This is UV (IR) dominant if $q_{\rm th,1}< 1 ~(>1)$ during the thermalization. In any case, slightly before $t_{\rm th,1}$,  the ignition rate is still faster than the Hubble parameter satisfying \Eq{condth}. 

In the time regime, $ \(\D t_{\rm ignition}\)^{-1}\gg H$, during the thermalization of $\chi_1$ \Eq{eq} is reached with $f_{\chi_1}<f_{\chi_1}^{\rm eq}$.\footnote{Strictly speaking, when $\G_{\rm th,1}$  is slower than $H$, that is slower than $\D t_{\rm ignition}^{-1}$,
the back reaction to $f_{\chi,1}$ due to the interaction \eq{decay} exists. 
It decreases $f_{\chi_1}$ at $t<t_{\rm th,1}$ compared to the discussed case neglecting this backreaction. 
The decrease is in a way satisfying the comoving number conservation $-\D(n_{\chi_1} a^3)= \D (n_{\chi_2} a^3)=\D (n_\f a^3)$ via \Eq{decay}.
Taking account of this effect should not change our conclusions because, in the end, we will get ${\chi_1}$ thermalized, with $f_{\chi_2}(p_{\chi_2}\sim T)\sim f_{\chi_1}(p_{\chi_1}\sim T)\sim f_{\chi_1}^{\rm eq}(p_{\chi_1}\sim T)$. During the whole process \Eq{conserv} is guaranteed.  }
Since $n_\f^{\rm burst} \sim g_{\chi_2}\int_{p_{\chi_2}\sim T} d^3p_{\chi_2} f_{\chi_2} \sim g_{\chi_2}\int_{p_{\chi_1}\sim T} d^3p_{\chi_1} f_{\chi_1}  \propto \frac{\G_{\rm th}}{H} T^3 \propto a^{-3+q_{\rm th,1}} $, 
 the comoving number density of $n_\f^{\rm burst}a^3$ increases in time. 
 At $t>t_{\rm th,1},$ it is frozen out as discussed in \Sec{4-1}.
 Therefore the dominant production happens at the cosmic temperature $T=T_{\rm th,1}$ which is the cosmic temperature that $\chi_1$ is fully thermalized, $f_{\chi_1}\approx f_{\chi_1}^{\rm eq}.$ At this moment, we get $f_{\chi_2}\sim f_{\chi_1}^{\rm eq}.$ 
Thus 
\beq
T_{\rm prod}\sim T_{\rm th,1}. 
\eeq
I numerically checked a similar behavior in the setup more-or-less close to this scenario with a simple modification, \Eq{initial}$\to f_{\chi_1}=\tanh (\Gamma_{\rm th,1}/H)\(e^{E_{\chi_1}/T}\mp 1 \)^{-1}$ with $a\propto t^{1/2}$.

Lastly, I comment on the thermalization rate of $\G_{\rm th,2}$ and $\G_{\rm th,\f}$.
$\G_{\rm th,2}$ for the momentum mode around $T$ should be slower than the $H$ at least at $T=T_{\rm th,1}$ since, otherwise, the $p_{\chi_2}\sim T$ modes of $\chi_2$ reach the equilibrium, and thus the resulting burst-produced $\f$ is suppressed (c.f. \Eq{conserv} is only for the reaction of \Eq{decay}, and \Sec{thermalization}). 
Similarly, $\G_{\rm th,\f}$ for the low momentum mode should be smaller than $H$ at the production. 
In addition, after the burst production, $\G_{\rm th,\f}$ for the low momentum mode may also be required to be smaller than the Hubble parameter because otherwise, the produced $\f$ is washed out. This is a usual assumption for the light DM. The consistency of the argument is checked by introducing the terms $-\Gamma_{\rm th,\f}(\hat f_\f[\hat p_\f]-\hat f_{\f}^{\rm eq}[\hat p_{\f}]),\AND -\Gamma_{\rm th,2}(\hat f_{\chi_2}[\hat p_{\chi_2}]-\hat f_{ \chi_2}^{\rm eq}[\hat p_{\chi_2}])$ in the Boltzmann equations \eq{fphi} and \eq{fpsi}, respectively. That said, I emphasize that the arguments may have exceptions due to the momentum dependence of the reaction and Bose enhancement. A more detailed model-dependent analysis by performing the Boltzmann equation will be desired.

\subsection{Model-building --case of ALP coupled to right-handed neutrinos--} 
\lac{model}
Let me roughly discuss possible models for the burst production of $\f$ DM. 
By assuming that $\f$ is an SM gauge singlet, $\chi_{1,2}$ should have the representation for the gauge group. In the case, $\chi_{1,2}$ has a non-trivial representation of the SM gauge group, the requirement $\G_{\rm th,2}\ll H$ can be only satisfied in the high-temperature regime, for instance, $T_R\gtrsim 10^{13-14}\GEV$ $\SU(2)$ gauge interactions are decoupled (e.g.~\cite{Hamada:2018epb}). In this case, we can use charged heavy beyond SM (BSM) particles to play the roles of $\chi_1$ in the burst production of $\f$. 
We will not consider this possibility but focus on the case that $\chi_{1,2}$ are also gauge singlets. 
The theoretical candidates may be the BSM singlet scalars, and fermions in various BSM scenarios (see also the footnote.~\ref{ft:1}), e.g.,  some supersymmetric partners, or right-handed neutrinos (RHN), the latter of which I will explain in more detail. 

The RHNs may exist to explain the smallness of the active neutrino masses by the seesaw mechanism~\cite{Yanagida:1979as,GellMann:1980vs,Glashow:1979nm,Mohapatra:1979ia} (see also \cite{Minkowski:1977sc} and a UV completion for charge quantization predicting the neutrino mass scale~\cite{Yin:2018qcs}) and to produce baryon asymmetry via leptogenesis~\cite{Fukugita:1986hr,Pilaftsis:1997dr,Buchmuller:1997yu,Akhmedov:1998qx,Asaka:2005pn} (see also \cite{Hamada:2016oft, Hamada:2018epb,Eijima:2019hey} in the effective theory with lepton flavor oscillation). 
The Lagrangian is given as 
\beq
\label{nuint}
 {\cal L}_{N} =i  \bar N_i \partial_\mu \gamma^\mu N_i -( \frac12 M_{ij} 
 \bar{N^c_i} N_j + y_{N,i\a} \bar{L}_\a \tl{H} \hat{P}_R N_i + {\rm h.c.}),
\eeq
where $L_\a$ is a left-handed lepton field in the chiral representation, $\tl H$ is the Higgs doublet field, $i,(\a)$ runs from $1$ to $n$ ($e,\m,\t$), and we take $M_{ij}=M_i \delta_{ij}$ and $M_i$ to be real without loss of generality. I do not restrict to the case $n=3\OR 2$ but take $n$ generic.

The thermalization of the RHN in the mass range of interest can be estimated as
\beq
\G_{N,i}^{\rm th}\simeq \gamma_{N} \sum_{\a}|y_{N,i\a}|^2 T
\eeq
with $\gamma_{N}\simeq0.01$ being the numerical result from Refs.~\cite{Besak:2012qm, Hernandez:2016kel} which includes $2 \leftrightarrow 2$ and $1 \leftrightarrow 2$ processes with SM particles as well as the Landau-Pomeranchuk-Migdal effects~\cite{Landau:1953um,Migdal:1956tc}. By comparing  $\G_{N,i}^{\rm th}$ with the Hubble parameter at the radiation dominated era, $H\simeq \sqrt{g_{\star} \pi^2 T^4/90M_{\rm pl}^2},$ one obtains the temperature that $N$ is thermalized 
\beq
\laq{thnt}
T_{{\rm th}, i}\simeq 7\TEV \(\frac{\sum_{\a}|y_{N,i\a}|}{10^{-6}}\)^2.
\eeq

The right-handed neutrino, $N_i$, can also couple to the light bosonic DM, especially the ALP, with a derivative coupling like
\beq
{\cal L}^{\rm int}_{\rm eff}\supset  \sum_{i\geq j} C_{N_iN_j} \frac{\partial_\mu \phi}{2f_\phi} \bar{N_i}\gamma_5\gamma^\mu {N_j}. 
\eeq
Moving to a mass basis by field redefinitions to remove $\f$ in the derivative, we obtain, 
\beq
{\cal L}_{\rm mass}= i  \sum_{i\geq j} C_{N_{i} N_j}(M_{i}+M_{j})\frac{\phi}{2f_\phi} \bar{N^c_i} \gamma_5 N_j.
\eeq
This interaction introduces the decay of 
\beq
N_1\to a N_{i\neq 1} \laq{Nreaction}
\eeq
where $N_1$ is the heaviest RHN. 
The ignition rate can be estimated as 
\beq
\(\D t_{\rm iginition}\)^{-1}\sim\sum_{i\neq 1}  C_{N_1,N_{i}}^2  \frac{T^3}{4\pi f_\f^2}.
\eeq
After this timescale, the ALP burst production occurs, stimulating all reactions \eq{Nreaction} via the Bose enhancement if the ALP couplings in the mass basis are not exponentially small.
We can easily find that the ignition rate is faster than the Hubble expansion rate at $T$ if 
\beq
f_\f\lesssim 2\times 10^9\GEV  \sqrt{\sum_{i\neq 1}  C_{N_1,N_{i}}^2}\sqrt{\frac{T}{100\GEV}}.
\eeq
From this, for the ALP with $f_\f\sim 10^{6-8}\GEV$ that is relevant to EBL hints and future reaches mentioned in the introduction, the process is very efficient. 
If $N_1$ has the highest thermalization temperature after reheating,
 \beq 
T_{R}\gg T_{{\rm th,}1} \gg T_{{\rm th,}i\neq1}~~ (\to\rm\Sec{thermalization})
\eeq 
this becomes the setup of \Sec{thermalization} with $T_{\rm prod}=T_{{\rm th}, 1}$, 
while if 
 \beq 
T_{{\rm th,}1} \gg T_{R}\gg T_{{\rm th,}i\neq1} ~~(\to\rm \Sec{reheating})
\eeq 
it becomes the setup of \Sec{reheating} with $T_{\rm prod}=T_{R}$.
In any case, $N_{i\neq 1}$ are gradually thermalized after the burst production via the reaction to all channels $N_1\to a N_{2},\cdots a N_n$. We can estimate the DM abundance with \Eq{abundance} by taking $g_{\chi_2}=2\times (n-1)$.
Via the active-neutrino Yukawa interactions, the comoving entropy carried by RHNs gets back to the SM much after the $\f$ burst production.

The hot component of $\f$ from the decay and inverse decay can be suppressed with the mild degeneracy of $M_1\sim M_{i\neq 1},$ which can lead to slightly larger $y_{N,ij}$ to explain the neutrino mass via the seesaw mechanism. 
Also, the hot component can be suppressed by considering relatively large $f_\f$ for the suppressed $\D t_{\rm decay}^{-1}$, or simply have DM light as discussed in \Sec{4-1}.\footnote{It is straightforward to apply the model to a hidden photon DM whose gauge coupling is not too large. Thanks to the equivalence theorem, we can consider $\f$ as the longitudinal mode of the photon with certain UV completions. 
% with gauge invariance, gauge anomaly cancellation, and Higgs mechanism. 
Model-independently, the ignition rate for $N_1$ decaying into the longitudinal mode and $N_{i\neq 1}$ does not change. The $N_1$ decay into the transverse mode is neglected because of the small gauge coupling. 
However, the discussion in the following, including the thermalization via photon coupling and decays into neutrinos, are only for the ALP.}

Since ALP is usually defined as an axion coupled to a pair of photons, I also check the thermalization of the ALP via the photon coupling.  
Thermalization rate for the $2 \to 2$ process involving an ALP and photon, e.g., ${e \bar{e} \to \g a} $, is roughly estimated $\G_{\rm th} \sim \a^3 T^3/f^2_\f$, which is several orders of magnitude smaller than the ignition rate. Here $\a\sim 1/128$ is the fine-structure constant.\footnote{This may be replaced by the one for SM $\SU(2)_L\OR \U(1)_Y$ coupling in the symmetric phase, which decreases the upper bound of $T$. With only $\U(1)_Y$ coupling, the discussion does not change much. } 
The thermalization does not occur for $T\lesssim T_{{\rm th},\phi}^{(\g)}\sim 0.3\TEV \frac{f_\f}{10^{7}\GEV}$. As long as this thermalization is inefficient at the burst production period at $T\sim T_{\rm prod} < T_{{\rm th},\phi}^{(\g)}$, the cold DM production happens. Even if the thermal relic of $\f$ exists at $T=T_{\rm prod}$ the burst production occurs (see the last part of \Sec{thermalization}). 
Interestingly, the hot component produced initially via the ALP-photon coupling is suppressed with $\eta<1$ due to the inverse decay $\f N_{i\neq 1}\to N_1$ at $T\sim M_{1}$, as numerically checked in the last panel of Fig.\ref{fig:2}.\footnote{This phenomenon may be also useful for suppressing the hot relics of the axion in the hadronic QCD axion window, an interesting eV range DM candidate~\cite{Chang:1993gm, Moroi:1998qs} (See also \cite{Chang:2018rso,Bar:2019ifz}) by introducing the BSM particles $\chi_{1,2}$ coupled to QCD axion.  Burst production after the freeze out of the axion-hadron interaction can be useful for the axion cold DM production if it can be made consistent with the big-bang nucleosynthesis.} 
Later, the ALP produced via the burst is kept intact because the dissipation rate via the interactions is suppressed by the tiny momentum as usual~\cite{Moroi:2014mqa} (see also another estimation by treating the ALP as an oscillating field~\cite{Nakayama:2021avl}). 
%This is 

A prediction of this scenario is that the ALP also decays into active neutrinos. The mass range can be reached by future cosmic neutrino background searches like PTOLEMY ~\cite{PTOLEMY:2018jst,McKeen:2018xyz,Chacko:2018uke}.

\section{Conclusions and discussion} 
\lac{5}
In this paper, I have shown that the thermal production of dark matter (DM) with a mass around eV may not result in the DM being as hot as has been considered. 
The coldness of the produced DM depends on the details of the reaction that produces the DM,  given that the TG bound favors the DM as a bosonic field in the mass range of eV. In a very short period, the bosonic DM may be burst-produced in much lower momentum modes than the cosmic temperature due to a Bose enhancement. 
Since the DM with the burst production naturally has the number density around that of the thermalized mother particles due to the back-reaction, the mass is predicted around $\EV$, the mass range of the conventional hot DM.
The cold component of the DM can remain until today if the cosmic expansion makes the burst reaction freeze out. One successful example has been discussed by focusing on the simple $1\leftrightarrow 2$ decay/inverse decay reaction
 without adopting the conventional approximations: (1) all the particles except for the DM are treated in thermal distribution and (2) neglect of the Bose-enhancement and Pauli-blocking factors. 
The resulting DM abundance predicts the mass in the range of 
\beq
m_{\rm DM}= \O\(1 - 100\)\EV.
\eeq 
 In summary, I claim that the eV mass range for the DM may still be special, and it should be theoretically well motivated. 

So far, I have used the simplest reaction to demonstrate my claim. 
There may be other examples of the light and cold bosonic DM production from hot plasma, e.g., via generic many to many scatterings, Bremsstrahlung emissions of hidden photons, etc., which are worth further studies. 

I also comment that in the whole discussion, I considered the possibility that various timescales have hierarchies, e.g., \Eq{cond}. 
Without the hierarchy, we may have less cold DM number and thus heavier DM mass, e.g., a few keV, for the abundance. Some examples were explained in footnotes \ref{ft:2} and \ref{ft:3}.  The DM in this scenario is colder than that of the usual thermally produced DM with the same mass. Thus the structure formation bounds for the DM can be relaxed.

\section*{Acknowledgments}
I thank Brian Batell for useful discussions on Boltzmann equations in another ongoing project.
The present work is supported by JSPS KAKENHI Grant Numbers  20H05851, 21K20364, 22K14029, and 22H01215.

\bibliography{main3}

%merlin.mbs apsrev4-1.bst 2010-07-25 4.21a (PWD, AO, DPC) hacked
%Control: key (0)
%Control: author (8) initials jnrlst
%Control: editor formatted (1) identically to author
%Control: production of article title (-1) disabled
%Control: page (0) single
%Control: year (1) truncated
%Control: production of eprint (0) enabled
\begin{thebibliography}{88}%
\makeatletter
\providecommand \@ifxundefined [1]{%
 \@ifx{#1\undefined}
}%
\providecommand \@ifnum [1]{%
 \ifnum #1\expandafter \@firstoftwo
 \else \expandafter \@secondoftwo
 \fi
}%
\providecommand \@ifx [1]{%
 \ifx #1\expandafter \@firstoftwo
 \else \expandafter \@secondoftwo
 \fi
}%
\providecommand \natexlab [1]{#1}%
\providecommand \enquote  [1]{``#1''}%
\providecommand \bibnamefont  [1]{#1}%
\providecommand \bibfnamefont [1]{#1}%
\providecommand \citenamefont [1]{#1}%
\providecommand \href@noop [0]{\@secondoftwo}%
\providecommand \href [0]{\begingroup \@sanitize@url \@href}%
\providecommand \@href[1]{\@@startlink{#1}\@@href}%
\providecommand \@@href[1]{\endgroup#1\@@endlink}%
\providecommand \@sanitize@url [0]{\catcode `\\12\catcode `\$12\catcode
  `\&12\catcode `\#12\catcode `\^12\catcode `\_12\catcode `\%12\relax}%
\providecommand \@@startlink[1]{}%
\providecommand \@@endlink[0]{}%
\providecommand \url  [0]{\begingroup\@sanitize@url \@url }%
\providecommand \@url [1]{\endgroup\@href {#1}{\urlprefix }}%
\providecommand \urlprefix  [0]{URL }%
\providecommand \Eprint [0]{\href }%
\providecommand \doibase [0]{http://dx.doi.org/}%
\providecommand \selectlanguage [0]{\@gobble}%
\providecommand \bibinfo  [0]{\@secondoftwo}%
\providecommand \bibfield  [0]{\@secondoftwo}%
\providecommand \translation [1]{[#1]}%
\providecommand \BibitemOpen [0]{}%
\providecommand \bibitemStop [0]{}%
\providecommand \bibitemNoStop [0]{.\EOS\space}%
\providecommand \EOS [0]{\spacefactor3000\relax}%
\providecommand \BibitemShut  [1]{\csname bibitem#1\endcsname}%
\let\auto@bib@innerbib\@empty
%</preamble>
\bibitem [{\citenamefont {Aghanim}\ \emph {et~al.}(2020)\citenamefont {Aghanim}
  \emph {et~al.}}]{Planck:2018vyg}%
  \BibitemOpen
  \bibfield  {author} {\bibinfo {author} {\bibfnamefont {N.}~\bibnamefont
  {Aghanim}} \emph {et~al.} (\bibinfo {collaboration} {Planck}),\ }\href
  {\doibase 10.1051/0004-6361/201833910} {\bibfield  {journal} {\bibinfo
  {journal} {Astron. Astrophys.}\ }\textbf {\bibinfo {volume} {641}},\ \bibinfo
  {pages} {A6} (\bibinfo {year} {2020})},\ \bibinfo {note} {[Erratum:
  Astron.Astrophys. 652, C4 (2021)]},\ \Eprint
  {http://arxiv.org/abs/1807.06209} {arXiv:1807.06209 [astro-ph.CO]}
  \BibitemShut {NoStop}%
\bibitem [{\citenamefont {Frenk}\ and\ \citenamefont
  {White}(2012)}]{Frenk:2012ph}%
  \BibitemOpen
  \bibfield  {author} {\bibinfo {author} {\bibfnamefont {C.~S.}\ \bibnamefont
  {Frenk}}\ and\ \bibinfo {author} {\bibfnamefont {S.~D.~M.}\ \bibnamefont
  {White}},\ }\href {\doibase 10.1002/andp.201200212} {\bibfield  {journal}
  {\bibinfo  {journal} {Annalen Phys.}\ }\textbf {\bibinfo {volume} {524}},\
  \bibinfo {pages} {507} (\bibinfo {year} {2012})},\ \Eprint
  {http://arxiv.org/abs/1210.0544} {arXiv:1210.0544 [astro-ph.CO]} \BibitemShut
  {NoStop}%
\bibitem [{\citenamefont {Viel}\ \emph {et~al.}(2005)\citenamefont {Viel},
  \citenamefont {Lesgourgues}, \citenamefont {Haehnelt}, \citenamefont
  {Matarrese},\ and\ \citenamefont {Riotto}}]{Viel:2005qj}%
  \BibitemOpen
  \bibfield  {author} {\bibinfo {author} {\bibfnamefont {M.}~\bibnamefont
  {Viel}}, \bibinfo {author} {\bibfnamefont {J.}~\bibnamefont {Lesgourgues}},
  \bibinfo {author} {\bibfnamefont {M.~G.}\ \bibnamefont {Haehnelt}}, \bibinfo
  {author} {\bibfnamefont {S.}~\bibnamefont {Matarrese}}, \ and\ \bibinfo
  {author} {\bibfnamefont {A.}~\bibnamefont {Riotto}},\ }\href {\doibase
  10.1103/PhysRevD.71.063534} {\bibfield  {journal} {\bibinfo  {journal} {Phys.
  Rev. D}\ }\textbf {\bibinfo {volume} {71}},\ \bibinfo {pages} {063534}
  (\bibinfo {year} {2005})},\ \Eprint {http://arxiv.org/abs/astro-ph/0501562}
  {arXiv:astro-ph/0501562} \BibitemShut {NoStop}%
\bibitem [{\citenamefont {Ir\v{s}i\v{c}}\ \emph {et~al.}(2017)\citenamefont
  {Ir\v{s}i\v{c}} \emph {et~al.}}]{Irsic:2017ixq}%
  \BibitemOpen
  \bibfield  {author} {\bibinfo {author} {\bibfnamefont {V.}~\bibnamefont
  {Ir\v{s}i\v{c}}} \emph {et~al.},\ }\href {\doibase
  10.1103/PhysRevD.96.023522} {\bibfield  {journal} {\bibinfo  {journal} {Phys.
  Rev. D}\ }\textbf {\bibinfo {volume} {96}},\ \bibinfo {pages} {023522}
  (\bibinfo {year} {2017})},\ \Eprint {http://arxiv.org/abs/1702.01764}
  {arXiv:1702.01764 [astro-ph.CO]} \BibitemShut {NoStop}%
\bibitem [{\citenamefont {Tremaine}\ and\ \citenamefont
  {Gunn}(1979)}]{Tremaine:1979we}%
  \BibitemOpen
  \bibfield  {author} {\bibinfo {author} {\bibfnamefont {S.}~\bibnamefont
  {Tremaine}}\ and\ \bibinfo {author} {\bibfnamefont {J.~E.}\ \bibnamefont
  {Gunn}},\ }\href {\doibase 10.1103/PhysRevLett.42.407} {\bibfield  {journal}
  {\bibinfo  {journal} {Phys. Rev. Lett.}\ }\textbf {\bibinfo {volume} {42}},\
  \bibinfo {pages} {407} (\bibinfo {year} {1979})}\BibitemShut {NoStop}%
\bibitem [{\citenamefont {Boyarsky}\ \emph {et~al.}(2009)\citenamefont
  {Boyarsky}, \citenamefont {Ruchayskiy},\ and\ \citenamefont
  {Iakubovskyi}}]{Boyarsky:2008ju}%
  \BibitemOpen
  \bibfield  {author} {\bibinfo {author} {\bibfnamefont {A.}~\bibnamefont
  {Boyarsky}}, \bibinfo {author} {\bibfnamefont {O.}~\bibnamefont
  {Ruchayskiy}}, \ and\ \bibinfo {author} {\bibfnamefont {D.}~\bibnamefont
  {Iakubovskyi}},\ }\href {\doibase 10.1088/1475-7516/2009/03/005} {\bibfield
  {journal} {\bibinfo  {journal} {JCAP}\ }\textbf {\bibinfo {volume} {03}},\
  \bibinfo {pages} {005} (\bibinfo {year} {2009})},\ \Eprint
  {http://arxiv.org/abs/0808.3902} {arXiv:0808.3902 [hep-ph]} \BibitemShut
  {NoStop}%
\bibitem [{\citenamefont {Randall}\ \emph {et~al.}(2017)\citenamefont
  {Randall}, \citenamefont {Scholtz},\ and\ \citenamefont
  {Unwin}}]{Randall:2016bqw}%
  \BibitemOpen
  \bibfield  {author} {\bibinfo {author} {\bibfnamefont {L.}~\bibnamefont
  {Randall}}, \bibinfo {author} {\bibfnamefont {J.}~\bibnamefont {Scholtz}}, \
  and\ \bibinfo {author} {\bibfnamefont {J.}~\bibnamefont {Unwin}},\ }\href
  {\doibase 10.1093/mnras/stx161} {\bibfield  {journal} {\bibinfo  {journal}
  {Mon. Not. Roy. Astron. Soc.}\ }\textbf {\bibinfo {volume} {467}},\ \bibinfo
  {pages} {1515} (\bibinfo {year} {2017})},\ \Eprint
  {http://arxiv.org/abs/1611.04590} {arXiv:1611.04590 [astro-ph.GA]}
  \BibitemShut {NoStop}%
\bibitem [{\citenamefont {Hall}\ \emph {et~al.}(2010)\citenamefont {Hall},
  \citenamefont {Jedamzik}, \citenamefont {March-Russell},\ and\ \citenamefont
  {West}}]{Hall:2009bx}%
  \BibitemOpen
  \bibfield  {author} {\bibinfo {author} {\bibfnamefont {L.~J.}\ \bibnamefont
  {Hall}}, \bibinfo {author} {\bibfnamefont {K.}~\bibnamefont {Jedamzik}},
  \bibinfo {author} {\bibfnamefont {J.}~\bibnamefont {March-Russell}}, \ and\
  \bibinfo {author} {\bibfnamefont {S.~M.}\ \bibnamefont {West}},\ }\href
  {\doibase 10.1007/JHEP03(2010)080} {\bibfield  {journal} {\bibinfo  {journal}
  {JHEP}\ }\textbf {\bibinfo {volume} {03}},\ \bibinfo {pages} {080} (\bibinfo
  {year} {2010})},\ \Eprint {http://arxiv.org/abs/0911.1120} {arXiv:0911.1120
  [hep-ph]} \BibitemShut {NoStop}%
\bibitem [{\citenamefont {Kamada}\ and\ \citenamefont
  {Yanagi}(2019)}]{Kamada:2019kpe}%
  \BibitemOpen
  \bibfield  {author} {\bibinfo {author} {\bibfnamefont {A.}~\bibnamefont
  {Kamada}}\ and\ \bibinfo {author} {\bibfnamefont {K.}~\bibnamefont
  {Yanagi}},\ }\href {\doibase 10.1088/1475-7516/2019/11/029} {\bibfield
  {journal} {\bibinfo  {journal} {JCAP}\ }\textbf {\bibinfo {volume} {11}},\
  \bibinfo {pages} {029} (\bibinfo {year} {2019})},\ \Eprint
  {http://arxiv.org/abs/1907.04558} {arXiv:1907.04558 [hep-ph]} \BibitemShut
  {NoStop}%
\bibitem [{\citenamefont {D'Eramo}\ and\ \citenamefont
  {Lenoci}(2021)}]{DEramo:2020gpr}%
  \BibitemOpen
  \bibfield  {author} {\bibinfo {author} {\bibfnamefont {F.}~\bibnamefont
  {D'Eramo}}\ and\ \bibinfo {author} {\bibfnamefont {A.}~\bibnamefont
  {Lenoci}},\ }\href {\doibase 10.1088/1475-7516/2021/10/045} {\bibfield
  {journal} {\bibinfo  {journal} {JCAP}\ }\textbf {\bibinfo {volume} {10}},\
  \bibinfo {pages} {045} (\bibinfo {year} {2021})},\ \Eprint
  {http://arxiv.org/abs/2012.01446} {arXiv:2012.01446 [hep-ph]} \BibitemShut
  {NoStop}%
\bibitem [{\citenamefont {Gong}\ \emph {et~al.}(2016)\citenamefont {Gong},
  \citenamefont {Cooray}, \citenamefont {Mitchell-Wynne}, \citenamefont {Chen},
  \citenamefont {Zemcov},\ and\ \citenamefont {Smidt}}]{Gong:2015hke}%
  \BibitemOpen
  \bibfield  {author} {\bibinfo {author} {\bibfnamefont {Y.}~\bibnamefont
  {Gong}}, \bibinfo {author} {\bibfnamefont {A.}~\bibnamefont {Cooray}},
  \bibinfo {author} {\bibfnamefont {K.}~\bibnamefont {Mitchell-Wynne}},
  \bibinfo {author} {\bibfnamefont {X.}~\bibnamefont {Chen}}, \bibinfo {author}
  {\bibfnamefont {M.}~\bibnamefont {Zemcov}}, \ and\ \bibinfo {author}
  {\bibfnamefont {J.}~\bibnamefont {Smidt}},\ }\href {\doibase
  10.3847/0004-637X/825/2/104} {\bibfield  {journal} {\bibinfo  {journal}
  {Astrophys. J.}\ }\textbf {\bibinfo {volume} {825}},\ \bibinfo {pages} {104}
  (\bibinfo {year} {2016})},\ \Eprint {http://arxiv.org/abs/1511.01577}
  {arXiv:1511.01577 [astro-ph.CO]} \BibitemShut {NoStop}%
\bibitem [{\citenamefont {Korochkin}\ \emph {et~al.}(2020)\citenamefont
  {Korochkin}, \citenamefont {Neronov},\ and\ \citenamefont
  {Semikoz}}]{Korochkin:2019qpe}%
  \BibitemOpen
  \bibfield  {author} {\bibinfo {author} {\bibfnamefont {A.}~\bibnamefont
  {Korochkin}}, \bibinfo {author} {\bibfnamefont {A.}~\bibnamefont {Neronov}},
  \ and\ \bibinfo {author} {\bibfnamefont {D.}~\bibnamefont {Semikoz}},\ }\href
  {\doibase 10.1088/1475-7516/2020/03/064} {\bibfield  {journal} {\bibinfo
  {journal} {JCAP}\ }\textbf {\bibinfo {volume} {03}},\ \bibinfo {pages} {064}
  (\bibinfo {year} {2020})},\ \Eprint {http://arxiv.org/abs/1911.13291}
  {arXiv:1911.13291 [hep-ph]} \BibitemShut {NoStop}%
\bibitem [{\citenamefont {Caputo}\ \emph {et~al.}(2021)\citenamefont {Caputo},
  \citenamefont {Vittino}, \citenamefont {Fornengo}, \citenamefont {Regis},\
  and\ \citenamefont {Taoso}}]{Caputo:2020msf}%
  \BibitemOpen
  \bibfield  {author} {\bibinfo {author} {\bibfnamefont {A.}~\bibnamefont
  {Caputo}}, \bibinfo {author} {\bibfnamefont {A.}~\bibnamefont {Vittino}},
  \bibinfo {author} {\bibfnamefont {N.}~\bibnamefont {Fornengo}}, \bibinfo
  {author} {\bibfnamefont {M.}~\bibnamefont {Regis}}, \ and\ \bibinfo {author}
  {\bibfnamefont {M.}~\bibnamefont {Taoso}},\ }\href {\doibase
  10.1088/1475-7516/2021/05/046} {\bibfield  {journal} {\bibinfo  {journal}
  {JCAP}\ }\textbf {\bibinfo {volume} {05}},\ \bibinfo {pages} {046} (\bibinfo
  {year} {2021})},\ \Eprint {http://arxiv.org/abs/2012.09179} {arXiv:2012.09179
  [astro-ph.CO]} \BibitemShut {NoStop}%
\bibitem [{\citenamefont {Bernal}\ \emph
  {et~al.}(2022{\natexlab{a}})\citenamefont {Bernal}, \citenamefont {Caputo},
  \citenamefont {Sato-Polito}, \citenamefont {Mirocha},\ and\ \citenamefont
  {Kamionkowski}}]{Bernal:2022xyi}%
  \BibitemOpen
  \bibfield  {author} {\bibinfo {author} {\bibfnamefont {J.~L.}\ \bibnamefont
  {Bernal}}, \bibinfo {author} {\bibfnamefont {A.}~\bibnamefont {Caputo}},
  \bibinfo {author} {\bibfnamefont {G.}~\bibnamefont {Sato-Polito}}, \bibinfo
  {author} {\bibfnamefont {J.}~\bibnamefont {Mirocha}}, \ and\ \bibinfo
  {author} {\bibfnamefont {M.}~\bibnamefont {Kamionkowski}},\ }\href@noop {} {\
   (\bibinfo {year} {2022}{\natexlab{a}})},\ \Eprint
  {http://arxiv.org/abs/2208.13794} {arXiv:2208.13794 [astro-ph.CO]}
  \BibitemShut {NoStop}%
\bibitem [{\citenamefont {Kohri}\ \emph {et~al.}(2017)\citenamefont {Kohri},
  \citenamefont {Moroi},\ and\ \citenamefont {Nakayama}}]{Kohri:2017oqn}%
  \BibitemOpen
  \bibfield  {author} {\bibinfo {author} {\bibfnamefont {K.}~\bibnamefont
  {Kohri}}, \bibinfo {author} {\bibfnamefont {T.}~\bibnamefont {Moroi}}, \ and\
  \bibinfo {author} {\bibfnamefont {K.}~\bibnamefont {Nakayama}},\ }\href
  {\doibase 10.1016/j.physletb.2017.07.026} {\bibfield  {journal} {\bibinfo
  {journal} {Phys. Lett. B}\ }\textbf {\bibinfo {volume} {772}},\ \bibinfo
  {pages} {628} (\bibinfo {year} {2017})},\ \Eprint
  {http://arxiv.org/abs/1706.04921} {arXiv:1706.04921 [astro-ph.CO]}
  \BibitemShut {NoStop}%
\bibitem [{\citenamefont {Kalashev}\ \emph {et~al.}(2019)\citenamefont
  {Kalashev}, \citenamefont {Kusenko},\ and\ \citenamefont
  {Vitagliano}}]{Kalashev:2018bra}%
  \BibitemOpen
  \bibfield  {author} {\bibinfo {author} {\bibfnamefont {O.~E.}\ \bibnamefont
  {Kalashev}}, \bibinfo {author} {\bibfnamefont {A.}~\bibnamefont {Kusenko}}, \
  and\ \bibinfo {author} {\bibfnamefont {E.}~\bibnamefont {Vitagliano}},\
  }\href {\doibase 10.1103/PhysRevD.99.023002} {\bibfield  {journal} {\bibinfo
  {journal} {Phys. Rev. D}\ }\textbf {\bibinfo {volume} {99}},\ \bibinfo
  {pages} {023002} (\bibinfo {year} {2019})},\ \Eprint
  {http://arxiv.org/abs/1808.05613} {arXiv:1808.05613 [hep-ph]} \BibitemShut
  {NoStop}%
\bibitem [{\citenamefont {Kashlinsky}\ \emph {et~al.}(2018)\citenamefont
  {Kashlinsky}, \citenamefont {Arendt}, \citenamefont {Atrio-Barandela},
  \citenamefont {Cappelluti}, \citenamefont {Ferrara},\ and\ \citenamefont
  {Hasinger}}]{Kashlinsky:2018mnu}%
  \BibitemOpen
  \bibfield  {author} {\bibinfo {author} {\bibfnamefont {A.}~\bibnamefont
  {Kashlinsky}}, \bibinfo {author} {\bibfnamefont {R.~G.}\ \bibnamefont
  {Arendt}}, \bibinfo {author} {\bibfnamefont {F.}~\bibnamefont
  {Atrio-Barandela}}, \bibinfo {author} {\bibfnamefont {N.}~\bibnamefont
  {Cappelluti}}, \bibinfo {author} {\bibfnamefont {A.}~\bibnamefont {Ferrara}},
  \ and\ \bibinfo {author} {\bibfnamefont {G.}~\bibnamefont {Hasinger}},\
  }\href {\doibase 10.1103/RevModPhys.90.025006} {\bibfield  {journal}
  {\bibinfo  {journal} {Rev. Mod. Phys.}\ }\textbf {\bibinfo {volume} {90}},\
  \bibinfo {pages} {025006} (\bibinfo {year} {2018})},\ \Eprint
  {http://arxiv.org/abs/1802.07774} {arXiv:1802.07774 [astro-ph.CO]}
  \BibitemShut {NoStop}%
\bibitem [{\citenamefont {Lauer}\ \emph {et~al.}(2022)\citenamefont {Lauer}
  \emph {et~al.}}]{Lauer:2022fgc}%
  \BibitemOpen
  \bibfield  {author} {\bibinfo {author} {\bibfnamefont {T.~R.}\ \bibnamefont
  {Lauer}} \emph {et~al.},\ }\href {\doibase 10.3847/2041-8213/ac573d}
  {\bibfield  {journal} {\bibinfo  {journal} {Astrophys. J. Lett.}\ }\textbf
  {\bibinfo {volume} {927}},\ \bibinfo {pages} {L8} (\bibinfo {year} {2022})},\
  \Eprint {http://arxiv.org/abs/2202.04273} {arXiv:2202.04273 [astro-ph.GA]}
  \BibitemShut {NoStop}%
\bibitem [{\citenamefont {Nakayama}\ and\ \citenamefont
  {Yin}(2022)}]{Nakayama:2022jza}%
  \BibitemOpen
  \bibfield  {author} {\bibinfo {author} {\bibfnamefont {K.}~\bibnamefont
  {Nakayama}}\ and\ \bibinfo {author} {\bibfnamefont {W.}~\bibnamefont {Yin}},\
  }\href {\doibase 10.1103/PhysRevD.106.103505} {\bibfield  {journal} {\bibinfo
   {journal} {Phys. Rev. D}\ }\textbf {\bibinfo {volume} {106}},\ \bibinfo
  {pages} {103505} (\bibinfo {year} {2022})},\ \Eprint
  {http://arxiv.org/abs/2205.01079} {arXiv:2205.01079 [hep-ph]} \BibitemShut
  {NoStop}%
\bibitem [{\citenamefont {Carenza}\ \emph {et~al.}(2023)\citenamefont
  {Carenza}, \citenamefont {Lucente},\ and\ \citenamefont
  {Vitagliano}}]{Carenza:2023qxh}%
  \BibitemOpen
  \bibfield  {author} {\bibinfo {author} {\bibfnamefont {P.}~\bibnamefont
  {Carenza}}, \bibinfo {author} {\bibfnamefont {G.}~\bibnamefont {Lucente}}, \
  and\ \bibinfo {author} {\bibfnamefont {E.}~\bibnamefont {Vitagliano}},\
  }\href@noop {} {\  (\bibinfo {year} {2023})},\ \Eprint
  {http://arxiv.org/abs/2301.06560} {arXiv:2301.06560 [hep-ph]} \BibitemShut
  {NoStop}%
\bibitem [{\citenamefont {Bernal}\ \emph
  {et~al.}(2022{\natexlab{b}})\citenamefont {Bernal}, \citenamefont
  {Sato-Polito},\ and\ \citenamefont {Kamionkowski}}]{Bernal:2022wsu}%
  \BibitemOpen
  \bibfield  {author} {\bibinfo {author} {\bibfnamefont {J.~L.}\ \bibnamefont
  {Bernal}}, \bibinfo {author} {\bibfnamefont {G.}~\bibnamefont {Sato-Polito}},
  \ and\ \bibinfo {author} {\bibfnamefont {M.}~\bibnamefont {Kamionkowski}},\
  }\href {\doibase 10.1103/PhysRevLett.129.231301} {\bibfield  {journal}
  {\bibinfo  {journal} {Phys. Rev. Lett.}\ }\textbf {\bibinfo {volume} {129}},\
  \bibinfo {pages} {231301} (\bibinfo {year} {2022}{\natexlab{b}})},\ \Eprint
  {http://arxiv.org/abs/2203.11236} {arXiv:2203.11236 [astro-ph.CO]}
  \BibitemShut {NoStop}%
\bibitem [{\citenamefont {Baryakhtar}\ \emph {et~al.}(2018)\citenamefont
  {Baryakhtar}, \citenamefont {Huang},\ and\ \citenamefont
  {Lasenby}}]{Baryakhtar:2018doz}%
  \BibitemOpen
  \bibfield  {author} {\bibinfo {author} {\bibfnamefont {M.}~\bibnamefont
  {Baryakhtar}}, \bibinfo {author} {\bibfnamefont {J.}~\bibnamefont {Huang}}, \
  and\ \bibinfo {author} {\bibfnamefont {R.}~\bibnamefont {Lasenby}},\ }\href
  {\doibase 10.1103/PhysRevD.98.035006} {\bibfield  {journal} {\bibinfo
  {journal} {Phys. Rev. D}\ }\textbf {\bibinfo {volume} {98}},\ \bibinfo
  {pages} {035006} (\bibinfo {year} {2018})},\ \Eprint
  {http://arxiv.org/abs/1803.11455} {arXiv:1803.11455 [hep-ph]} \BibitemShut
  {NoStop}%
\bibitem [{\citenamefont {Bessho}\ \emph {et~al.}(2022)\citenamefont {Bessho},
  \citenamefont {Ikeda},\ and\ \citenamefont {Yin}}]{Bessho:2022yyu}%
  \BibitemOpen
  \bibfield  {author} {\bibinfo {author} {\bibfnamefont {T.}~\bibnamefont
  {Bessho}}, \bibinfo {author} {\bibfnamefont {Y.}~\bibnamefont {Ikeda}}, \
  and\ \bibinfo {author} {\bibfnamefont {W.}~\bibnamefont {Yin}},\ }\href
  {\doibase 10.1103/PhysRevD.106.095025} {\bibfield  {journal} {\bibinfo
  {journal} {Phys. Rev. D}\ }\textbf {\bibinfo {volume} {106}},\ \bibinfo
  {pages} {095025} (\bibinfo {year} {2022})},\ \Eprint
  {http://arxiv.org/abs/2208.05975} {arXiv:2208.05975 [hep-ph]} \BibitemShut
  {NoStop}%
\bibitem [{\citenamefont {Shirasaki}(2021)}]{Shirasaki:2021yrp}%
  \BibitemOpen
  \bibfield  {author} {\bibinfo {author} {\bibfnamefont {M.}~\bibnamefont
  {Shirasaki}},\ }\href {\doibase 10.1103/PhysRevD.103.103014} {\bibfield
  {journal} {\bibinfo  {journal} {Phys. Rev. D}\ }\textbf {\bibinfo {volume}
  {103}},\ \bibinfo {pages} {103014} (\bibinfo {year} {2021})},\ \Eprint
  {http://arxiv.org/abs/2102.00580} {arXiv:2102.00580 [astro-ph.CO]}
  \BibitemShut {NoStop}%
\bibitem [{\citenamefont {Irastorza}\ \emph {et~al.}(2011)\citenamefont
  {Irastorza} \emph {et~al.}}]{Irastorza:2011gs}%
  \BibitemOpen
  \bibfield  {author} {\bibinfo {author} {\bibfnamefont {I.~G.}\ \bibnamefont
  {Irastorza}} \emph {et~al.},\ }\href {\doibase 10.1088/1475-7516/2011/06/013}
  {\bibfield  {journal} {\bibinfo  {journal} {JCAP}\ }\textbf {\bibinfo
  {volume} {06}},\ \bibinfo {pages} {013} (\bibinfo {year} {2011})},\ \Eprint
  {http://arxiv.org/abs/1103.5334} {arXiv:1103.5334 [hep-ex]} \BibitemShut
  {NoStop}%
\bibitem [{\citenamefont {Armengaud}\ \emph {et~al.}(2014)\citenamefont
  {Armengaud} \emph {et~al.}}]{Armengaud:2014gea}%
  \BibitemOpen
  \bibfield  {author} {\bibinfo {author} {\bibfnamefont {E.}~\bibnamefont
  {Armengaud}} \emph {et~al.},\ }\href {\doibase 10.1088/1748-0221/9/05/T05002}
  {\bibfield  {journal} {\bibinfo  {journal} {JINST}\ }\textbf {\bibinfo
  {volume} {9}},\ \bibinfo {pages} {T05002} (\bibinfo {year} {2014})},\ \Eprint
  {http://arxiv.org/abs/1401.3233} {arXiv:1401.3233 [physics.ins-det]}
  \BibitemShut {NoStop}%
\bibitem [{\citenamefont {Armengaud}\ \emph {et~al.}(2019)\citenamefont
  {Armengaud} \emph {et~al.}}]{Armengaud:2019uso}%
  \BibitemOpen
  \bibfield  {author} {\bibinfo {author} {\bibfnamefont {E.}~\bibnamefont
  {Armengaud}} \emph {et~al.} (\bibinfo {collaboration} {IAXO}),\ }\href
  {\doibase 10.1088/1475-7516/2019/06/047} {\bibfield  {journal} {\bibinfo
  {journal} {JCAP}\ }\textbf {\bibinfo {volume} {06}},\ \bibinfo {pages} {047}
  (\bibinfo {year} {2019})},\ \Eprint {http://arxiv.org/abs/1904.09155}
  {arXiv:1904.09155 [hep-ph]} \BibitemShut {NoStop}%
\bibitem [{\citenamefont {Abeln}\ \emph {et~al.}(2021)\citenamefont {Abeln}
  \emph {et~al.}}]{Abeln:2020ywv}%
  \BibitemOpen
  \bibfield  {author} {\bibinfo {author} {\bibfnamefont {A.}~\bibnamefont
  {Abeln}} \emph {et~al.} (\bibinfo {collaboration} {IAXO}),\ }\href {\doibase
  10.1007/JHEP05(2021)137} {\bibfield  {journal} {\bibinfo  {journal} {JHEP}\
  }\textbf {\bibinfo {volume} {05}},\ \bibinfo {pages} {137} (\bibinfo {year}
  {2021})},\ \Eprint {http://arxiv.org/abs/2010.12076} {arXiv:2010.12076
  [physics.ins-det]} \BibitemShut {NoStop}%
\bibitem [{\citenamefont {Homma}\ \emph {et~al.}(2023)\citenamefont {Homma},
  \citenamefont {Ishibashi}, \citenamefont {Kirita},\ and\ \citenamefont
  {Hasada}}]{Homma:2022ktv}%
  \BibitemOpen
  \bibfield  {author} {\bibinfo {author} {\bibfnamefont {K.}~\bibnamefont
  {Homma}}, \bibinfo {author} {\bibfnamefont {F.}~\bibnamefont {Ishibashi}},
  \bibinfo {author} {\bibfnamefont {Y.}~\bibnamefont {Kirita}}, \ and\ \bibinfo
  {author} {\bibfnamefont {T.}~\bibnamefont {Hasada}},\ }\href {\doibase
  10.3390/universe9010020} {\bibfield  {journal} {\bibinfo  {journal}
  {Universe}\ }\textbf {\bibinfo {volume} {9}},\ \bibinfo {pages} {20}
  (\bibinfo {year} {2023})},\ \Eprint {http://arxiv.org/abs/2212.13012}
  {arXiv:2212.13012 [hep-ph]} \BibitemShut {NoStop}%
\bibitem [{\citenamefont {Daido}\ \emph {et~al.}(2017)\citenamefont {Daido},
  \citenamefont {Takahashi},\ and\ \citenamefont {Yin}}]{Daido:2017wwb}%
  \BibitemOpen
  \bibfield  {author} {\bibinfo {author} {\bibfnamefont {R.}~\bibnamefont
  {Daido}}, \bibinfo {author} {\bibfnamefont {F.}~\bibnamefont {Takahashi}}, \
  and\ \bibinfo {author} {\bibfnamefont {W.}~\bibnamefont {Yin}},\ }\href
  {\doibase 10.1088/1475-7516/2017/05/044} {\bibfield  {journal} {\bibinfo
  {journal} {JCAP}\ }\textbf {\bibinfo {volume} {05}},\ \bibinfo {pages} {044}
  (\bibinfo {year} {2017})},\ \Eprint {http://arxiv.org/abs/1702.03284}
  {arXiv:1702.03284 [hep-ph]} \BibitemShut {NoStop}%
\bibitem [{\citenamefont {Daido}\ \emph {et~al.}(2018)\citenamefont {Daido},
  \citenamefont {Takahashi},\ and\ \citenamefont {Yin}}]{Daido:2017tbr}%
  \BibitemOpen
  \bibfield  {author} {\bibinfo {author} {\bibfnamefont {R.}~\bibnamefont
  {Daido}}, \bibinfo {author} {\bibfnamefont {F.}~\bibnamefont {Takahashi}}, \
  and\ \bibinfo {author} {\bibfnamefont {W.}~\bibnamefont {Yin}},\ }\href
  {\doibase 10.1007/JHEP02(2018)104} {\bibfield  {journal} {\bibinfo  {journal}
  {JHEP}\ }\textbf {\bibinfo {volume} {02}},\ \bibinfo {pages} {104} (\bibinfo
  {year} {2018})},\ \Eprint {http://arxiv.org/abs/1710.11107} {arXiv:1710.11107
  [hep-ph]} \BibitemShut {NoStop}%
\bibitem [{\citenamefont {Graham}\ and\ \citenamefont
  {Scherlis}(2018)}]{Graham:2018jyp}%
  \BibitemOpen
  \bibfield  {author} {\bibinfo {author} {\bibfnamefont {P.~W.}\ \bibnamefont
  {Graham}}\ and\ \bibinfo {author} {\bibfnamefont {A.}~\bibnamefont
  {Scherlis}},\ }\href {\doibase 10.1103/PhysRevD.98.035017} {\bibfield
  {journal} {\bibinfo  {journal} {Phys. Rev. D}\ }\textbf {\bibinfo {volume}
  {98}},\ \bibinfo {pages} {035017} (\bibinfo {year} {2018})},\ \Eprint
  {http://arxiv.org/abs/1805.07362} {arXiv:1805.07362 [hep-ph]} \BibitemShut
  {NoStop}%
\bibitem [{\citenamefont {Takahashi}\ \emph {et~al.}(2018)\citenamefont
  {Takahashi}, \citenamefont {Yin},\ and\ \citenamefont {Guth}}]{Guth:2018hsa}%
  \BibitemOpen
  \bibfield  {author} {\bibinfo {author} {\bibfnamefont {F.}~\bibnamefont
  {Takahashi}}, \bibinfo {author} {\bibfnamefont {W.}~\bibnamefont {Yin}}, \
  and\ \bibinfo {author} {\bibfnamefont {A.~H.}\ \bibnamefont {Guth}},\ }\href
  {\doibase 10.1103/PhysRevD.98.015042} {\bibfield  {journal} {\bibinfo
  {journal} {Phys. Rev. D}\ }\textbf {\bibinfo {volume} {98}},\ \bibinfo
  {pages} {015042} (\bibinfo {year} {2018})},\ \Eprint
  {http://arxiv.org/abs/1805.08763} {arXiv:1805.08763 [hep-ph]} \BibitemShut
  {NoStop}%
\bibitem [{\citenamefont {Ho}\ \emph {et~al.}(2019)\citenamefont {Ho},
  \citenamefont {Takahashi},\ and\ \citenamefont {Yin}}]{Ho:2019ayl}%
  \BibitemOpen
  \bibfield  {author} {\bibinfo {author} {\bibfnamefont {S.-Y.}\ \bibnamefont
  {Ho}}, \bibinfo {author} {\bibfnamefont {F.}~\bibnamefont {Takahashi}}, \
  and\ \bibinfo {author} {\bibfnamefont {W.}~\bibnamefont {Yin}},\ }\href
  {\doibase 10.1007/JHEP04(2019)149} {\bibfield  {journal} {\bibinfo  {journal}
  {JHEP}\ }\textbf {\bibinfo {volume} {04}},\ \bibinfo {pages} {149} (\bibinfo
  {year} {2019})},\ \Eprint {http://arxiv.org/abs/1901.01240} {arXiv:1901.01240
  [hep-ph]} \BibitemShut {NoStop}%
\bibitem [{\citenamefont {Graham}\ \emph {et~al.}(2016)\citenamefont {Graham},
  \citenamefont {Mardon},\ and\ \citenamefont {Rajendran}}]{Graham:2015rva}%
  \BibitemOpen
  \bibfield  {author} {\bibinfo {author} {\bibfnamefont {P.~W.}\ \bibnamefont
  {Graham}}, \bibinfo {author} {\bibfnamefont {J.}~\bibnamefont {Mardon}}, \
  and\ \bibinfo {author} {\bibfnamefont {S.}~\bibnamefont {Rajendran}},\ }\href
  {\doibase 10.1103/PhysRevD.93.103520} {\bibfield  {journal} {\bibinfo
  {journal} {Phys. Rev. D}\ }\textbf {\bibinfo {volume} {93}},\ \bibinfo
  {pages} {103520} (\bibinfo {year} {2016})},\ \Eprint
  {http://arxiv.org/abs/1504.02102} {arXiv:1504.02102 [hep-ph]} \BibitemShut
  {NoStop}%
\bibitem [{\citenamefont {Ema}\ \emph {et~al.}(2019)\citenamefont {Ema},
  \citenamefont {Nakayama},\ and\ \citenamefont {Tang}}]{Ema:2019yrd}%
  \BibitemOpen
  \bibfield  {author} {\bibinfo {author} {\bibfnamefont {Y.}~\bibnamefont
  {Ema}}, \bibinfo {author} {\bibfnamefont {K.}~\bibnamefont {Nakayama}}, \
  and\ \bibinfo {author} {\bibfnamefont {Y.}~\bibnamefont {Tang}},\ }\href
  {\doibase 10.1007/JHEP07(2019)060} {\bibfield  {journal} {\bibinfo  {journal}
  {JHEP}\ }\textbf {\bibinfo {volume} {07}},\ \bibinfo {pages} {060} (\bibinfo
  {year} {2019})},\ \Eprint {http://arxiv.org/abs/1903.10973} {arXiv:1903.10973
  [hep-ph]} \BibitemShut {NoStop}%
\bibitem [{\citenamefont {Moroi}\ and\ \citenamefont
  {Yin}(2021{\natexlab{a}})}]{Moroi:2020has}%
  \BibitemOpen
  \bibfield  {author} {\bibinfo {author} {\bibfnamefont {T.}~\bibnamefont
  {Moroi}}\ and\ \bibinfo {author} {\bibfnamefont {W.}~\bibnamefont {Yin}},\
  }\href {\doibase 10.1007/JHEP03(2021)301} {\bibfield  {journal} {\bibinfo
  {journal} {JHEP}\ }\textbf {\bibinfo {volume} {03}},\ \bibinfo {pages} {301}
  (\bibinfo {year} {2021}{\natexlab{a}})},\ \Eprint
  {http://arxiv.org/abs/2011.09475} {arXiv:2011.09475 [hep-ph]} \BibitemShut
  {NoStop}%
\bibitem [{\citenamefont {Moroi}\ and\ \citenamefont
  {Yin}(2021{\natexlab{b}})}]{Moroi:2020bkq}%
  \BibitemOpen
  \bibfield  {author} {\bibinfo {author} {\bibfnamefont {T.}~\bibnamefont
  {Moroi}}\ and\ \bibinfo {author} {\bibfnamefont {W.}~\bibnamefont {Yin}},\
  }\href {\doibase 10.1007/JHEP03(2021)296} {\bibfield  {journal} {\bibinfo
  {journal} {JHEP}\ }\textbf {\bibinfo {volume} {03}},\ \bibinfo {pages} {296}
  (\bibinfo {year} {2021}{\natexlab{b}})},\ \Eprint
  {http://arxiv.org/abs/2011.12285} {arXiv:2011.12285 [hep-ph]} \BibitemShut
  {NoStop}%
\bibitem [{\citenamefont {Arias}\ \emph {et~al.}(2012)\citenamefont {Arias},
  \citenamefont {Cadamuro}, \citenamefont {Goodsell}, \citenamefont {Jaeckel},
  \citenamefont {Redondo},\ and\ \citenamefont {Ringwald}}]{Arias:2012az}%
  \BibitemOpen
  \bibfield  {author} {\bibinfo {author} {\bibfnamefont {P.}~\bibnamefont
  {Arias}}, \bibinfo {author} {\bibfnamefont {D.}~\bibnamefont {Cadamuro}},
  \bibinfo {author} {\bibfnamefont {M.}~\bibnamefont {Goodsell}}, \bibinfo
  {author} {\bibfnamefont {J.}~\bibnamefont {Jaeckel}}, \bibinfo {author}
  {\bibfnamefont {J.}~\bibnamefont {Redondo}}, \ and\ \bibinfo {author}
  {\bibfnamefont {A.}~\bibnamefont {Ringwald}},\ }\href {\doibase
  10.1088/1475-7516/2012/06/013} {\bibfield  {journal} {\bibinfo  {journal}
  {JCAP}\ }\textbf {\bibinfo {volume} {06}},\ \bibinfo {pages} {013} (\bibinfo
  {year} {2012})},\ \Eprint {http://arxiv.org/abs/1201.5902} {arXiv:1201.5902
  [hep-ph]} \BibitemShut {NoStop}%
\bibitem [{\citenamefont {Nakagawa}\ \emph {et~al.}(2020)\citenamefont
  {Nakagawa}, \citenamefont {Takahashi},\ and\ \citenamefont
  {Yin}}]{Nakagawa:2020eeg}%
  \BibitemOpen
  \bibfield  {author} {\bibinfo {author} {\bibfnamefont {S.}~\bibnamefont
  {Nakagawa}}, \bibinfo {author} {\bibfnamefont {F.}~\bibnamefont {Takahashi}},
  \ and\ \bibinfo {author} {\bibfnamefont {W.}~\bibnamefont {Yin}},\ }\href
  {\doibase 10.1088/1475-7516/2020/05/004} {\bibfield  {journal} {\bibinfo
  {journal} {JCAP}\ }\textbf {\bibinfo {volume} {05}},\ \bibinfo {pages} {004}
  (\bibinfo {year} {2020})},\ \Eprint {http://arxiv.org/abs/2002.12195}
  {arXiv:2002.12195 [hep-ph]} \BibitemShut {NoStop}%
\bibitem [{\citenamefont {Marsh}\ and\ \citenamefont
  {Yin}(2021)}]{Marsh:2019bjr}%
  \BibitemOpen
  \bibfield  {author} {\bibinfo {author} {\bibfnamefont {D.~J.~E.}\
  \bibnamefont {Marsh}}\ and\ \bibinfo {author} {\bibfnamefont
  {W.}~\bibnamefont {Yin}},\ }\href {\doibase 10.1007/JHEP01(2021)169}
  {\bibfield  {journal} {\bibinfo  {journal} {JHEP}\ }\textbf {\bibinfo
  {volume} {01}},\ \bibinfo {pages} {169} (\bibinfo {year} {2021})},\ \Eprint
  {http://arxiv.org/abs/1912.08188} {arXiv:1912.08188 [hep-ph]} \BibitemShut
  {NoStop}%
\bibitem [{\citenamefont {Sakurai}\ and\ \citenamefont
  {Yin}(2022{\natexlab{a}})}]{Sakurai:2021ipp}%
  \BibitemOpen
  \bibfield  {author} {\bibinfo {author} {\bibfnamefont {K.}~\bibnamefont
  {Sakurai}}\ and\ \bibinfo {author} {\bibfnamefont {W.}~\bibnamefont {Yin}},\
  }\href {\doibase 10.1007/JHEP04(2022)113} {\bibfield  {journal} {\bibinfo
  {journal} {JHEP}\ }\textbf {\bibinfo {volume} {04}},\ \bibinfo {pages} {113}
  (\bibinfo {year} {2022}{\natexlab{a}})},\ \Eprint
  {http://arxiv.org/abs/2111.03653} {arXiv:2111.03653 [hep-ph]} \BibitemShut
  {NoStop}%
\bibitem [{\citenamefont {Sakurai}\ and\ \citenamefont
  {Yin}(2022{\natexlab{b}})}]{Sakurai:2022cki}%
  \BibitemOpen
  \bibfield  {author} {\bibinfo {author} {\bibfnamefont {K.}~\bibnamefont
  {Sakurai}}\ and\ \bibinfo {author} {\bibfnamefont {W.}~\bibnamefont {Yin}},\
  }\href@noop {} {\  (\bibinfo {year} {2022}{\natexlab{b}})},\ \Eprint
  {http://arxiv.org/abs/2204.01739} {arXiv:2204.01739 [hep-ph]} \BibitemShut
  {NoStop}%
\bibitem [{\citenamefont {Haghighat}\ \emph {et~al.}(2022)\citenamefont
  {Haghighat}, \citenamefont {Mohammadi~Najafabadi}, \citenamefont {Sakurai},\
  and\ \citenamefont {Yin}}]{Haghighat:2022qyh}%
  \BibitemOpen
  \bibfield  {author} {\bibinfo {author} {\bibfnamefont {G.}~\bibnamefont
  {Haghighat}}, \bibinfo {author} {\bibfnamefont {M.}~\bibnamefont
  {Mohammadi~Najafabadi}}, \bibinfo {author} {\bibfnamefont {K.}~\bibnamefont
  {Sakurai}}, \ and\ \bibinfo {author} {\bibfnamefont {W.}~\bibnamefont
  {Yin}},\ }\href@noop {} {\  (\bibinfo {year} {2022})},\ \Eprint
  {http://arxiv.org/abs/2209.07565} {arXiv:2209.07565 [hep-ph]} \BibitemShut
  {NoStop}%
\bibitem [{\citenamefont {Kolb}\ and\ \citenamefont
  {Turner}(1990)}]{Kolb:1990vq}%
  \BibitemOpen
  \bibfield  {author} {\bibinfo {author} {\bibfnamefont {E.~W.}\ \bibnamefont
  {Kolb}}\ and\ \bibinfo {author} {\bibfnamefont {M.~S.}\ \bibnamefont
  {Turner}},\ }\href {\doibase 10.1201/9780429492860} {\emph {\bibinfo {title}
  {{The Early Universe}}}},\ Vol.~\bibinfo {volume} {69}\ (\bibinfo {year}
  {1990})\BibitemShut {NoStop}%
\bibitem [{\citenamefont {Li}\ \emph {et~al.}(2021)\citenamefont {Li},
  \citenamefont {Moroi}, \citenamefont {Nakayama},\ and\ \citenamefont
  {Yin}}]{Li:2021fao}%
  \BibitemOpen
  \bibfield  {author} {\bibinfo {author} {\bibfnamefont {Q.}~\bibnamefont
  {Li}}, \bibinfo {author} {\bibfnamefont {T.}~\bibnamefont {Moroi}}, \bibinfo
  {author} {\bibfnamefont {K.}~\bibnamefont {Nakayama}}, \ and\ \bibinfo
  {author} {\bibfnamefont {W.}~\bibnamefont {Yin}},\ }\href {\doibase
  10.1007/JHEP09(2021)179} {\bibfield  {journal} {\bibinfo  {journal} {JHEP}\
  }\textbf {\bibinfo {volume} {09}},\ \bibinfo {pages} {179} (\bibinfo {year}
  {2021})},\ \Eprint {http://arxiv.org/abs/2105.13358} {arXiv:2105.13358
  [hep-ph]} \BibitemShut {NoStop}%
\bibitem [{\citenamefont {Diamanti}\ \emph {et~al.}(2017)\citenamefont
  {Diamanti}, \citenamefont {Ando}, \citenamefont {Gariazzo}, \citenamefont
  {Mena},\ and\ \citenamefont {Weniger}}]{Diamanti:2017xfo}%
  \BibitemOpen
  \bibfield  {author} {\bibinfo {author} {\bibfnamefont {R.}~\bibnamefont
  {Diamanti}}, \bibinfo {author} {\bibfnamefont {S.}~\bibnamefont {Ando}},
  \bibinfo {author} {\bibfnamefont {S.}~\bibnamefont {Gariazzo}}, \bibinfo
  {author} {\bibfnamefont {O.}~\bibnamefont {Mena}}, \ and\ \bibinfo {author}
  {\bibfnamefont {C.}~\bibnamefont {Weniger}},\ }\href {\doibase
  10.1088/1475-7516/2017/06/008} {\bibfield  {journal} {\bibinfo  {journal}
  {JCAP}\ }\textbf {\bibinfo {volume} {06}},\ \bibinfo {pages} {008} (\bibinfo
  {year} {2017})},\ \Eprint {http://arxiv.org/abs/1701.03128} {arXiv:1701.03128
  [astro-ph.CO]} \BibitemShut {NoStop}%
\bibitem [{\citenamefont {Di~Luzio}\ \emph {et~al.}(2017)\citenamefont
  {Di~Luzio}, \citenamefont {Nardi},\ and\ \citenamefont
  {Ubaldi}}]{DiLuzio:2017tjx}%
  \BibitemOpen
  \bibfield  {author} {\bibinfo {author} {\bibfnamefont {L.}~\bibnamefont
  {Di~Luzio}}, \bibinfo {author} {\bibfnamefont {E.}~\bibnamefont {Nardi}}, \
  and\ \bibinfo {author} {\bibfnamefont {L.}~\bibnamefont {Ubaldi}},\ }\href
  {\doibase 10.1103/PhysRevLett.119.011801} {\bibfield  {journal} {\bibinfo
  {journal} {Phys. Rev. Lett.}\ }\textbf {\bibinfo {volume} {119}},\ \bibinfo
  {pages} {011801} (\bibinfo {year} {2017})},\ \Eprint
  {http://arxiv.org/abs/1704.01122} {arXiv:1704.01122 [hep-ph]} \BibitemShut
  {NoStop}%
\bibitem [{\citenamefont {Lee}\ and\ \citenamefont {Yin}(2019)}]{Lee:2018yak}%
  \BibitemOpen
  \bibfield  {author} {\bibinfo {author} {\bibfnamefont {H.-S.}\ \bibnamefont
  {Lee}}\ and\ \bibinfo {author} {\bibfnamefont {W.}~\bibnamefont {Yin}},\
  }\href {\doibase 10.1103/PhysRevD.99.015041} {\bibfield  {journal} {\bibinfo
  {journal} {Phys. Rev. D}\ }\textbf {\bibinfo {volume} {99}},\ \bibinfo
  {pages} {015041} (\bibinfo {year} {2019})},\ \Eprint
  {http://arxiv.org/abs/1811.04039} {arXiv:1811.04039 [hep-ph]} \BibitemShut
  {NoStop}%
\bibitem [{\citenamefont {Ardu}\ \emph {et~al.}(2020)\citenamefont {Ardu},
  \citenamefont {Di~Luzio}, \citenamefont {Landini}, \citenamefont {Strumia},
  \citenamefont {Teresi},\ and\ \citenamefont {Wang}}]{Ardu:2020qmo}%
  \BibitemOpen
  \bibfield  {author} {\bibinfo {author} {\bibfnamefont {M.}~\bibnamefont
  {Ardu}}, \bibinfo {author} {\bibfnamefont {L.}~\bibnamefont {Di~Luzio}},
  \bibinfo {author} {\bibfnamefont {G.}~\bibnamefont {Landini}}, \bibinfo
  {author} {\bibfnamefont {A.}~\bibnamefont {Strumia}}, \bibinfo {author}
  {\bibfnamefont {D.}~\bibnamefont {Teresi}}, \ and\ \bibinfo {author}
  {\bibfnamefont {J.-W.}\ \bibnamefont {Wang}},\ }\href {\doibase
  10.1007/JHEP11(2020)090} {\bibfield  {journal} {\bibinfo  {journal} {JHEP}\
  }\textbf {\bibinfo {volume} {11}},\ \bibinfo {pages} {090} (\bibinfo {year}
  {2020})},\ \Eprint {http://arxiv.org/abs/2007.12663} {arXiv:2007.12663
  [hep-ph]} \BibitemShut {NoStop}%
\bibitem [{\citenamefont {Yin}(2020)}]{Yin:2020dfn}%
  \BibitemOpen
  \bibfield  {author} {\bibinfo {author} {\bibfnamefont {W.}~\bibnamefont
  {Yin}},\ }\href {\doibase 10.1007/JHEP10(2020)032} {\bibfield  {journal}
  {\bibinfo  {journal} {JHEP}\ }\textbf {\bibinfo {volume} {10}},\ \bibinfo
  {pages} {032} (\bibinfo {year} {2020})},\ \Eprint
  {http://arxiv.org/abs/2007.13320} {arXiv:2007.13320 [hep-ph]} \BibitemShut
  {NoStop}%
\bibitem [{\citenamefont {Jaeckel}\ and\ \citenamefont
  {Yin}(2023)}]{Jaeckel:2022osh}%
  \BibitemOpen
  \bibfield  {author} {\bibinfo {author} {\bibfnamefont {J.}~\bibnamefont
  {Jaeckel}}\ and\ \bibinfo {author} {\bibfnamefont {W.}~\bibnamefont {Yin}},\
  }\href {\doibase 10.1103/PhysRevD.107.015001} {\bibfield  {journal} {\bibinfo
   {journal} {Phys. Rev. D}\ }\textbf {\bibinfo {volume} {107}},\ \bibinfo
  {pages} {015001} (\bibinfo {year} {2023})},\ \Eprint
  {http://arxiv.org/abs/2206.06376} {arXiv:2206.06376 [hep-ph]} \BibitemShut
  {NoStop}%
\bibitem [{\citenamefont {Agrawal}\ and\ \citenamefont
  {Howe}(2018)}]{Agrawal:2017ksf}%
  \BibitemOpen
  \bibfield  {author} {\bibinfo {author} {\bibfnamefont {P.}~\bibnamefont
  {Agrawal}}\ and\ \bibinfo {author} {\bibfnamefont {K.}~\bibnamefont {Howe}},\
  }\href {\doibase 10.1007/JHEP12(2018)029} {\bibfield  {journal} {\bibinfo
  {journal} {JHEP}\ }\textbf {\bibinfo {volume} {12}},\ \bibinfo {pages} {029}
  (\bibinfo {year} {2018})},\ \Eprint {http://arxiv.org/abs/1710.04213}
  {arXiv:1710.04213 [hep-ph]} \BibitemShut {NoStop}%
\bibitem [{\citenamefont {Takahashi}\ and\ \citenamefont
  {Yin}(2019)}]{Takahashi:2019qmh}%
  \BibitemOpen
  \bibfield  {author} {\bibinfo {author} {\bibfnamefont {F.}~\bibnamefont
  {Takahashi}}\ and\ \bibinfo {author} {\bibfnamefont {W.}~\bibnamefont
  {Yin}},\ }\href {\doibase 10.1007/JHEP07(2019)095} {\bibfield  {journal}
  {\bibinfo  {journal} {JHEP}\ }\textbf {\bibinfo {volume} {07}},\ \bibinfo
  {pages} {095} (\bibinfo {year} {2019})},\ \Eprint
  {http://arxiv.org/abs/1903.00462} {arXiv:1903.00462 [hep-ph]} \BibitemShut
  {NoStop}%
\bibitem [{\citenamefont {Takahashi}\ \emph {et~al.}(2021)\citenamefont
  {Takahashi}, \citenamefont {Yamada},\ and\ \citenamefont
  {Yin}}]{Takahashi:2020uio}%
  \BibitemOpen
  \bibfield  {author} {\bibinfo {author} {\bibfnamefont {F.}~\bibnamefont
  {Takahashi}}, \bibinfo {author} {\bibfnamefont {M.}~\bibnamefont {Yamada}}, \
  and\ \bibinfo {author} {\bibfnamefont {W.}~\bibnamefont {Yin}},\ }\href
  {\doibase 10.1007/JHEP01(2021)152} {\bibfield  {journal} {\bibinfo  {journal}
  {JHEP}\ }\textbf {\bibinfo {volume} {01}},\ \bibinfo {pages} {152} (\bibinfo
  {year} {2021})},\ \Eprint {http://arxiv.org/abs/2007.10311} {arXiv:2007.10311
  [hep-ph]} \BibitemShut {NoStop}%
\bibitem [{\citenamefont {Takahashi}\ and\ \citenamefont
  {Yin}(2021)}]{Takahashi:2021tff}%
  \BibitemOpen
  \bibfield  {author} {\bibinfo {author} {\bibfnamefont {F.}~\bibnamefont
  {Takahashi}}\ and\ \bibinfo {author} {\bibfnamefont {W.}~\bibnamefont
  {Yin}},\ }\href {\doibase 10.1088/1475-7516/2021/10/057} {\bibfield
  {journal} {\bibinfo  {journal} {JCAP}\ }\textbf {\bibinfo {volume} {10}},\
  \bibinfo {pages} {057} (\bibinfo {year} {2021})},\ \Eprint
  {http://arxiv.org/abs/2105.10493} {arXiv:2105.10493 [hep-ph]} \BibitemShut
  {NoStop}%
\bibitem [{\citenamefont {Azatov}\ \emph
  {et~al.}(2021{\natexlab{a}})\citenamefont {Azatov}, \citenamefont
  {Vanvlasselaer},\ and\ \citenamefont {Yin}}]{Azatov:2021irb}%
  \BibitemOpen
  \bibfield  {author} {\bibinfo {author} {\bibfnamefont {A.}~\bibnamefont
  {Azatov}}, \bibinfo {author} {\bibfnamefont {M.}~\bibnamefont
  {Vanvlasselaer}}, \ and\ \bibinfo {author} {\bibfnamefont {W.}~\bibnamefont
  {Yin}},\ }\href {\doibase 10.1007/JHEP10(2021)043} {\bibfield  {journal}
  {\bibinfo  {journal} {JHEP}\ }\textbf {\bibinfo {volume} {10}},\ \bibinfo
  {pages} {043} (\bibinfo {year} {2021}{\natexlab{a}})},\ \Eprint
  {http://arxiv.org/abs/2106.14913} {arXiv:2106.14913 [hep-ph]} \BibitemShut
  {NoStop}%
\bibitem [{\citenamefont {Baldes}\ \emph {et~al.}(2021)\citenamefont {Baldes},
  \citenamefont {Blasi}, \citenamefont {Mariotti}, \citenamefont {Sevrin},\
  and\ \citenamefont {Turbang}}]{Baldes:2021vyz}%
  \BibitemOpen
  \bibfield  {author} {\bibinfo {author} {\bibfnamefont {I.}~\bibnamefont
  {Baldes}}, \bibinfo {author} {\bibfnamefont {S.}~\bibnamefont {Blasi}},
  \bibinfo {author} {\bibfnamefont {A.}~\bibnamefont {Mariotti}}, \bibinfo
  {author} {\bibfnamefont {A.}~\bibnamefont {Sevrin}}, \ and\ \bibinfo {author}
  {\bibfnamefont {K.}~\bibnamefont {Turbang}},\ }\href {\doibase
  10.1103/PhysRevD.104.115029} {\bibfield  {journal} {\bibinfo  {journal}
  {Phys. Rev. D}\ }\textbf {\bibinfo {volume} {104}},\ \bibinfo {pages}
  {115029} (\bibinfo {year} {2021})},\ \Eprint
  {http://arxiv.org/abs/2106.15602} {arXiv:2106.15602 [hep-ph]} \BibitemShut
  {NoStop}%
\bibitem [{\citenamefont {Azatov}\ \emph {et~al.}(2022)\citenamefont {Azatov},
  \citenamefont {Barni}, \citenamefont {Chakraborty}, \citenamefont
  {Vanvlasselaer},\ and\ \citenamefont {Yin}}]{Azatov:2022tii}%
  \BibitemOpen
  \bibfield  {author} {\bibinfo {author} {\bibfnamefont {A.}~\bibnamefont
  {Azatov}}, \bibinfo {author} {\bibfnamefont {G.}~\bibnamefont {Barni}},
  \bibinfo {author} {\bibfnamefont {S.}~\bibnamefont {Chakraborty}}, \bibinfo
  {author} {\bibfnamefont {M.}~\bibnamefont {Vanvlasselaer}}, \ and\ \bibinfo
  {author} {\bibfnamefont {W.}~\bibnamefont {Yin}},\ }\href {\doibase
  10.1007/JHEP10(2022)017} {\bibfield  {journal} {\bibinfo  {journal} {JHEP}\
  }\textbf {\bibinfo {volume} {10}},\ \bibinfo {pages} {017} (\bibinfo {year}
  {2022})},\ \Eprint {http://arxiv.org/abs/2207.02230} {arXiv:2207.02230
  [hep-ph]} \BibitemShut {NoStop}%
\bibitem [{\citenamefont {Azatov}\ and\ \citenamefont
  {Vanvlasselaer}(2021)}]{Azatov:2020ufh}%
  \BibitemOpen
  \bibfield  {author} {\bibinfo {author} {\bibfnamefont {A.}~\bibnamefont
  {Azatov}}\ and\ \bibinfo {author} {\bibfnamefont {M.}~\bibnamefont
  {Vanvlasselaer}},\ }\href {\doibase 10.1088/1475-7516/2021/01/058} {\bibfield
   {journal} {\bibinfo  {journal} {JCAP}\ }\textbf {\bibinfo {volume} {01}},\
  \bibinfo {pages} {058} (\bibinfo {year} {2021})},\ \Eprint
  {http://arxiv.org/abs/2010.02590} {arXiv:2010.02590 [hep-ph]} \BibitemShut
  {NoStop}%
\bibitem [{\citenamefont {Azatov}\ \emph
  {et~al.}(2021{\natexlab{b}})\citenamefont {Azatov}, \citenamefont
  {Vanvlasselaer},\ and\ \citenamefont {Yin}}]{Azatov:2021ifm}%
  \BibitemOpen
  \bibfield  {author} {\bibinfo {author} {\bibfnamefont {A.}~\bibnamefont
  {Azatov}}, \bibinfo {author} {\bibfnamefont {M.}~\bibnamefont
  {Vanvlasselaer}}, \ and\ \bibinfo {author} {\bibfnamefont {W.}~\bibnamefont
  {Yin}},\ }\href {\doibase 10.1007/JHEP03(2021)288} {\bibfield  {journal}
  {\bibinfo  {journal} {JHEP}\ }\textbf {\bibinfo {volume} {03}},\ \bibinfo
  {pages} {288} (\bibinfo {year} {2021}{\natexlab{b}})},\ \Eprint
  {http://arxiv.org/abs/2101.05721} {arXiv:2101.05721 [hep-ph]} \BibitemShut
  {NoStop}%
\bibitem [{\citenamefont {Hamada}\ \emph {et~al.}(2018)\citenamefont {Hamada},
  \citenamefont {Kitano},\ and\ \citenamefont {Yin}}]{Hamada:2018epb}%
  \BibitemOpen
  \bibfield  {author} {\bibinfo {author} {\bibfnamefont {Y.}~\bibnamefont
  {Hamada}}, \bibinfo {author} {\bibfnamefont {R.}~\bibnamefont {Kitano}}, \
  and\ \bibinfo {author} {\bibfnamefont {W.}~\bibnamefont {Yin}},\ }\href
  {\doibase 10.1007/JHEP10(2018)178} {\bibfield  {journal} {\bibinfo  {journal}
  {JHEP}\ }\textbf {\bibinfo {volume} {10}},\ \bibinfo {pages} {178} (\bibinfo
  {year} {2018})},\ \Eprint {http://arxiv.org/abs/1807.06582} {arXiv:1807.06582
  [hep-ph]} \BibitemShut {NoStop}%
\bibitem [{\citenamefont {Yanagida}(1979)}]{Yanagida:1979as}%
  \BibitemOpen
  \bibfield  {author} {\bibinfo {author} {\bibfnamefont {T.}~\bibnamefont
  {Yanagida}},\ }\href@noop {} {\bibfield  {journal} {\bibinfo  {journal}
  {Conf. Proc. C}\ }\textbf {\bibinfo {volume} {7902131}},\ \bibinfo {pages}
  {95} (\bibinfo {year} {1979})}\BibitemShut {NoStop}%
\bibitem [{\citenamefont {Gell-Mann}\ \emph {et~al.}(1979)\citenamefont
  {Gell-Mann}, \citenamefont {Ramond},\ and\ \citenamefont
  {Slansky}}]{GellMann:1980vs}%
  \BibitemOpen
  \bibfield  {author} {\bibinfo {author} {\bibfnamefont {M.}~\bibnamefont
  {Gell-Mann}}, \bibinfo {author} {\bibfnamefont {P.}~\bibnamefont {Ramond}}, \
  and\ \bibinfo {author} {\bibfnamefont {R.}~\bibnamefont {Slansky}},\
  }\href@noop {} {\bibfield  {journal} {\bibinfo  {journal} {Conf. Proc. C}\
  }\textbf {\bibinfo {volume} {790927}},\ \bibinfo {pages} {315} (\bibinfo
  {year} {1979})},\ \Eprint {http://arxiv.org/abs/1306.4669} {arXiv:1306.4669
  [hep-th]} \BibitemShut {NoStop}%
\bibitem [{\citenamefont {Glashow}(1980)}]{Glashow:1979nm}%
  \BibitemOpen
  \bibfield  {author} {\bibinfo {author} {\bibfnamefont {S.~L.}\ \bibnamefont
  {Glashow}},\ }\href {\doibase 10.1007/978-1-4684-7197-7_15} {\bibfield
  {journal} {\bibinfo  {journal} {NATO Sci. Ser. B}\ }\textbf {\bibinfo
  {volume} {61}},\ \bibinfo {pages} {687} (\bibinfo {year} {1980})}\BibitemShut
  {NoStop}%
\bibitem [{\citenamefont {Mohapatra}\ and\ \citenamefont
  {Senjanovic}(1980)}]{Mohapatra:1979ia}%
  \BibitemOpen
  \bibfield  {author} {\bibinfo {author} {\bibfnamefont {R.~N.}\ \bibnamefont
  {Mohapatra}}\ and\ \bibinfo {author} {\bibfnamefont {G.}~\bibnamefont
  {Senjanovic}},\ }\href {\doibase 10.1103/PhysRevLett.44.912} {\bibfield
  {journal} {\bibinfo  {journal} {Phys. Rev. Lett.}\ }\textbf {\bibinfo
  {volume} {44}},\ \bibinfo {pages} {912} (\bibinfo {year} {1980})}\BibitemShut
  {NoStop}%
\bibitem [{\citenamefont {Minkowski}(1977)}]{Minkowski:1977sc}%
  \BibitemOpen
  \bibfield  {author} {\bibinfo {author} {\bibfnamefont {P.}~\bibnamefont
  {Minkowski}},\ }\href {\doibase 10.1016/0370-2693(77)90435-X} {\bibfield
  {journal} {\bibinfo  {journal} {Phys. Lett. B}\ }\textbf {\bibinfo {volume}
  {67}},\ \bibinfo {pages} {421} (\bibinfo {year} {1977})}\BibitemShut
  {NoStop}%
\bibitem [{\citenamefont {Yin}(2018)}]{Yin:2018qcs}%
  \BibitemOpen
  \bibfield  {author} {\bibinfo {author} {\bibfnamefont {W.}~\bibnamefont
  {Yin}},\ }\href {\doibase 10.1016/j.physletb.2018.09.023} {\bibfield
  {journal} {\bibinfo  {journal} {Phys. Lett. B}\ }\textbf {\bibinfo {volume}
  {785}},\ \bibinfo {pages} {585} (\bibinfo {year} {2018})},\ \Eprint
  {http://arxiv.org/abs/1808.00440} {arXiv:1808.00440 [hep-ph]} \BibitemShut
  {NoStop}%
\bibitem [{\citenamefont {Fukugita}\ and\ \citenamefont
  {Yanagida}(1986)}]{Fukugita:1986hr}%
  \BibitemOpen
  \bibfield  {author} {\bibinfo {author} {\bibfnamefont {M.}~\bibnamefont
  {Fukugita}}\ and\ \bibinfo {author} {\bibfnamefont {T.}~\bibnamefont
  {Yanagida}},\ }\href {\doibase 10.1016/0370-2693(86)91126-3} {\bibfield
  {journal} {\bibinfo  {journal} {Phys. Lett. B}\ }\textbf {\bibinfo {volume}
  {174}},\ \bibinfo {pages} {45} (\bibinfo {year} {1986})}\BibitemShut
  {NoStop}%
\bibitem [{\citenamefont {Pilaftsis}(1997)}]{Pilaftsis:1997dr}%
  \BibitemOpen
  \bibfield  {author} {\bibinfo {author} {\bibfnamefont {A.}~\bibnamefont
  {Pilaftsis}},\ }\href {\doibase 10.1016/S0550-3213(97)00469-0} {\bibfield
  {journal} {\bibinfo  {journal} {Nucl. Phys. B}\ }\textbf {\bibinfo {volume}
  {504}},\ \bibinfo {pages} {61} (\bibinfo {year} {1997})},\ \Eprint
  {http://arxiv.org/abs/hep-ph/9702393} {arXiv:hep-ph/9702393} \BibitemShut
  {NoStop}%
\bibitem [{\citenamefont {Buchmuller}\ and\ \citenamefont
  {Plumacher}(1998)}]{Buchmuller:1997yu}%
  \BibitemOpen
  \bibfield  {author} {\bibinfo {author} {\bibfnamefont {W.}~\bibnamefont
  {Buchmuller}}\ and\ \bibinfo {author} {\bibfnamefont {M.}~\bibnamefont
  {Plumacher}},\ }\href {\doibase 10.1016/S0370-2693(97)01548-7} {\bibfield
  {journal} {\bibinfo  {journal} {Phys. Lett. B}\ }\textbf {\bibinfo {volume}
  {431}},\ \bibinfo {pages} {354} (\bibinfo {year} {1998})},\ \Eprint
  {http://arxiv.org/abs/hep-ph/9710460} {arXiv:hep-ph/9710460} \BibitemShut
  {NoStop}%
\bibitem [{\citenamefont {Akhmedov}\ \emph {et~al.}(1998)\citenamefont
  {Akhmedov}, \citenamefont {Rubakov},\ and\ \citenamefont
  {Smirnov}}]{Akhmedov:1998qx}%
  \BibitemOpen
  \bibfield  {author} {\bibinfo {author} {\bibfnamefont {E.~K.}\ \bibnamefont
  {Akhmedov}}, \bibinfo {author} {\bibfnamefont {V.~A.}\ \bibnamefont
  {Rubakov}}, \ and\ \bibinfo {author} {\bibfnamefont {A.~Y.}\ \bibnamefont
  {Smirnov}},\ }\href {\doibase 10.1103/PhysRevLett.81.1359} {\bibfield
  {journal} {\bibinfo  {journal} {Phys. Rev. Lett.}\ }\textbf {\bibinfo
  {volume} {81}},\ \bibinfo {pages} {1359} (\bibinfo {year} {1998})},\ \Eprint
  {http://arxiv.org/abs/hep-ph/9803255} {arXiv:hep-ph/9803255} \BibitemShut
  {NoStop}%
\bibitem [{\citenamefont {Asaka}\ and\ \citenamefont
  {Shaposhnikov}(2005)}]{Asaka:2005pn}%
  \BibitemOpen
  \bibfield  {author} {\bibinfo {author} {\bibfnamefont {T.}~\bibnamefont
  {Asaka}}\ and\ \bibinfo {author} {\bibfnamefont {M.}~\bibnamefont
  {Shaposhnikov}},\ }\href {\doibase 10.1016/j.physletb.2005.06.020} {\bibfield
   {journal} {\bibinfo  {journal} {Phys. Lett. B}\ }\textbf {\bibinfo {volume}
  {620}},\ \bibinfo {pages} {17} (\bibinfo {year} {2005})},\ \Eprint
  {http://arxiv.org/abs/hep-ph/0505013} {arXiv:hep-ph/0505013} \BibitemShut
  {NoStop}%
\bibitem [{\citenamefont {Hamada}\ and\ \citenamefont
  {Kitano}(2016)}]{Hamada:2016oft}%
  \BibitemOpen
  \bibfield  {author} {\bibinfo {author} {\bibfnamefont {Y.}~\bibnamefont
  {Hamada}}\ and\ \bibinfo {author} {\bibfnamefont {R.}~\bibnamefont
  {Kitano}},\ }\href {\doibase 10.1007/JHEP11(2016)010} {\bibfield  {journal}
  {\bibinfo  {journal} {JHEP}\ }\textbf {\bibinfo {volume} {11}},\ \bibinfo
  {pages} {010} (\bibinfo {year} {2016})},\ \Eprint
  {http://arxiv.org/abs/1609.05028} {arXiv:1609.05028 [hep-ph]} \BibitemShut
  {NoStop}%
\bibitem [{\citenamefont {Eijima}\ \emph {et~al.}(2020)\citenamefont {Eijima},
  \citenamefont {Kitano},\ and\ \citenamefont {Yin}}]{Eijima:2019hey}%
  \BibitemOpen
  \bibfield  {author} {\bibinfo {author} {\bibfnamefont {S.}~\bibnamefont
  {Eijima}}, \bibinfo {author} {\bibfnamefont {R.}~\bibnamefont {Kitano}}, \
  and\ \bibinfo {author} {\bibfnamefont {W.}~\bibnamefont {Yin}},\ }\href
  {\doibase 10.1088/1475-7516/2020/03/048} {\bibfield  {journal} {\bibinfo
  {journal} {JCAP}\ }\textbf {\bibinfo {volume} {03}},\ \bibinfo {pages} {048}
  (\bibinfo {year} {2020})},\ \Eprint {http://arxiv.org/abs/1908.11864}
  {arXiv:1908.11864 [hep-ph]} \BibitemShut {NoStop}%
\bibitem [{\citenamefont {Besak}\ and\ \citenamefont
  {Bodeker}(2012)}]{Besak:2012qm}%
  \BibitemOpen
  \bibfield  {author} {\bibinfo {author} {\bibfnamefont {D.}~\bibnamefont
  {Besak}}\ and\ \bibinfo {author} {\bibfnamefont {D.}~\bibnamefont
  {Bodeker}},\ }\href {\doibase 10.1088/1475-7516/2012/03/029} {\bibfield
  {journal} {\bibinfo  {journal} {JCAP}\ }\textbf {\bibinfo {volume} {03}},\
  \bibinfo {pages} {029} (\bibinfo {year} {2012})},\ \Eprint
  {http://arxiv.org/abs/1202.1288} {arXiv:1202.1288 [hep-ph]} \BibitemShut
  {NoStop}%
\bibitem [{\citenamefont {Hern\'andez}\ \emph {et~al.}(2016)\citenamefont
  {Hern\'andez}, \citenamefont {Kekic}, \citenamefont {L\'opez-Pav\'on},
  \citenamefont {Racker},\ and\ \citenamefont {Salvado}}]{Hernandez:2016kel}%
  \BibitemOpen
  \bibfield  {author} {\bibinfo {author} {\bibfnamefont {P.}~\bibnamefont
  {Hern\'andez}}, \bibinfo {author} {\bibfnamefont {M.}~\bibnamefont {Kekic}},
  \bibinfo {author} {\bibfnamefont {J.}~\bibnamefont {L\'opez-Pav\'on}},
  \bibinfo {author} {\bibfnamefont {J.}~\bibnamefont {Racker}}, \ and\ \bibinfo
  {author} {\bibfnamefont {J.}~\bibnamefont {Salvado}},\ }\href {\doibase
  10.1007/JHEP08(2016)157} {\bibfield  {journal} {\bibinfo  {journal} {JHEP}\
  }\textbf {\bibinfo {volume} {08}},\ \bibinfo {pages} {157} (\bibinfo {year}
  {2016})},\ \Eprint {http://arxiv.org/abs/1606.06719} {arXiv:1606.06719
  [hep-ph]} \BibitemShut {NoStop}%
\bibitem [{\citenamefont {Landau}\ and\ \citenamefont
  {Pomeranchuk}(1953)}]{Landau:1953um}%
  \BibitemOpen
  \bibfield  {author} {\bibinfo {author} {\bibfnamefont {L.~D.}\ \bibnamefont
  {Landau}}\ and\ \bibinfo {author} {\bibfnamefont {I.}~\bibnamefont
  {Pomeranchuk}},\ }\href@noop {} {\bibfield  {journal} {\bibinfo  {journal}
  {Dokl. Akad. Nauk Ser. Fiz.}\ }\textbf {\bibinfo {volume} {92}},\ \bibinfo
  {pages} {535} (\bibinfo {year} {1953})}\BibitemShut {NoStop}%
\bibitem [{\citenamefont {Migdal}(1956)}]{Migdal:1956tc}%
  \BibitemOpen
  \bibfield  {author} {\bibinfo {author} {\bibfnamefont {A.~B.}\ \bibnamefont
  {Migdal}},\ }\href {\doibase 10.1103/PhysRev.103.1811} {\bibfield  {journal}
  {\bibinfo  {journal} {Phys. Rev.}\ }\textbf {\bibinfo {volume} {103}},\
  \bibinfo {pages} {1811} (\bibinfo {year} {1956})}\BibitemShut {NoStop}%
\bibitem [{\citenamefont {Chang}\ and\ \citenamefont
  {Choi}(1993)}]{Chang:1993gm}%
  \BibitemOpen
  \bibfield  {author} {\bibinfo {author} {\bibfnamefont {S.}~\bibnamefont
  {Chang}}\ and\ \bibinfo {author} {\bibfnamefont {K.}~\bibnamefont {Choi}},\
  }\href {\doibase 10.1016/0370-2693(93)90656-3} {\bibfield  {journal}
  {\bibinfo  {journal} {Phys. Lett. B}\ }\textbf {\bibinfo {volume} {316}},\
  \bibinfo {pages} {51} (\bibinfo {year} {1993})},\ \Eprint
  {http://arxiv.org/abs/hep-ph/9306216} {arXiv:hep-ph/9306216} \BibitemShut
  {NoStop}%
\bibitem [{\citenamefont {Moroi}\ and\ \citenamefont
  {Murayama}(1998)}]{Moroi:1998qs}%
  \BibitemOpen
  \bibfield  {author} {\bibinfo {author} {\bibfnamefont {T.}~\bibnamefont
  {Moroi}}\ and\ \bibinfo {author} {\bibfnamefont {H.}~\bibnamefont
  {Murayama}},\ }\href {\doibase 10.1016/S0370-2693(98)01091-0} {\bibfield
  {journal} {\bibinfo  {journal} {Phys. Lett. B}\ }\textbf {\bibinfo {volume}
  {440}},\ \bibinfo {pages} {69} (\bibinfo {year} {1998})},\ \Eprint
  {http://arxiv.org/abs/hep-ph/9804291} {arXiv:hep-ph/9804291} \BibitemShut
  {NoStop}%
\bibitem [{\citenamefont {Chang}\ \emph {et~al.}(2018)\citenamefont {Chang},
  \citenamefont {Essig},\ and\ \citenamefont {McDermott}}]{Chang:2018rso}%
  \BibitemOpen
  \bibfield  {author} {\bibinfo {author} {\bibfnamefont {J.~H.}\ \bibnamefont
  {Chang}}, \bibinfo {author} {\bibfnamefont {R.}~\bibnamefont {Essig}}, \ and\
  \bibinfo {author} {\bibfnamefont {S.~D.}\ \bibnamefont {McDermott}},\ }\href
  {\doibase 10.1007/JHEP09(2018)051} {\bibfield  {journal} {\bibinfo  {journal}
  {JHEP}\ }\textbf {\bibinfo {volume} {09}},\ \bibinfo {pages} {051} (\bibinfo
  {year} {2018})},\ \Eprint {http://arxiv.org/abs/1803.00993} {arXiv:1803.00993
  [hep-ph]} \BibitemShut {NoStop}%
\bibitem [{\citenamefont {Bar}\ \emph {et~al.}(2020)\citenamefont {Bar},
  \citenamefont {Blum},\ and\ \citenamefont {D'Amico}}]{Bar:2019ifz}%
  \BibitemOpen
  \bibfield  {author} {\bibinfo {author} {\bibfnamefont {N.}~\bibnamefont
  {Bar}}, \bibinfo {author} {\bibfnamefont {K.}~\bibnamefont {Blum}}, \ and\
  \bibinfo {author} {\bibfnamefont {G.}~\bibnamefont {D'Amico}},\ }\href
  {\doibase 10.1103/PhysRevD.101.123025} {\bibfield  {journal} {\bibinfo
  {journal} {Phys. Rev. D}\ }\textbf {\bibinfo {volume} {101}},\ \bibinfo
  {pages} {123025} (\bibinfo {year} {2020})},\ \Eprint
  {http://arxiv.org/abs/1907.05020} {arXiv:1907.05020 [hep-ph]} \BibitemShut
  {NoStop}%
\bibitem [{\citenamefont {Moroi}\ \emph {et~al.}(2014)\citenamefont {Moroi},
  \citenamefont {Mukaida}, \citenamefont {Nakayama},\ and\ \citenamefont
  {Takimoto}}]{Moroi:2014mqa}%
  \BibitemOpen
  \bibfield  {author} {\bibinfo {author} {\bibfnamefont {T.}~\bibnamefont
  {Moroi}}, \bibinfo {author} {\bibfnamefont {K.}~\bibnamefont {Mukaida}},
  \bibinfo {author} {\bibfnamefont {K.}~\bibnamefont {Nakayama}}, \ and\
  \bibinfo {author} {\bibfnamefont {M.}~\bibnamefont {Takimoto}},\ }\href
  {\doibase 10.1007/JHEP11(2014)151} {\bibfield  {journal} {\bibinfo  {journal}
  {JHEP}\ }\textbf {\bibinfo {volume} {11}},\ \bibinfo {pages} {151} (\bibinfo
  {year} {2014})},\ \Eprint {http://arxiv.org/abs/1407.7465} {arXiv:1407.7465
  [hep-ph]} \BibitemShut {NoStop}%
\bibitem [{\citenamefont {Nakayama}\ and\ \citenamefont
  {Yin}(2021)}]{Nakayama:2021avl}%
  \BibitemOpen
  \bibfield  {author} {\bibinfo {author} {\bibfnamefont {K.}~\bibnamefont
  {Nakayama}}\ and\ \bibinfo {author} {\bibfnamefont {W.}~\bibnamefont {Yin}},\
  }\href {\doibase 10.1007/JHEP10(2021)026} {\bibfield  {journal} {\bibinfo
  {journal} {JHEP}\ }\textbf {\bibinfo {volume} {10}},\ \bibinfo {pages} {026}
  (\bibinfo {year} {2021})},\ \Eprint {http://arxiv.org/abs/2105.14549}
  {arXiv:2105.14549 [hep-ph]} \BibitemShut {NoStop}%
\bibitem [{\citenamefont {Baracchini}\ \emph {et~al.}(2018)\citenamefont
  {Baracchini} \emph {et~al.}}]{PTOLEMY:2018jst}%
  \BibitemOpen
  \bibfield  {author} {\bibinfo {author} {\bibfnamefont {E.}~\bibnamefont
  {Baracchini}} \emph {et~al.} (\bibinfo {collaboration} {PTOLEMY}),\
  }\href@noop {} {\  (\bibinfo {year} {2018})},\ \Eprint
  {http://arxiv.org/abs/1808.01892} {arXiv:1808.01892 [physics.ins-det]}
  \BibitemShut {NoStop}%
\bibitem [{\citenamefont {McKeen}(2019)}]{McKeen:2018xyz}%
  \BibitemOpen
  \bibfield  {author} {\bibinfo {author} {\bibfnamefont {D.}~\bibnamefont
  {McKeen}},\ }\href {\doibase 10.1103/PhysRevD.100.015028} {\bibfield
  {journal} {\bibinfo  {journal} {Phys. Rev. D}\ }\textbf {\bibinfo {volume}
  {100}},\ \bibinfo {pages} {015028} (\bibinfo {year} {2019})},\ \Eprint
  {http://arxiv.org/abs/1812.08178} {arXiv:1812.08178 [hep-ph]} \BibitemShut
  {NoStop}%
\bibitem [{\citenamefont {Chacko}\ \emph {et~al.}(2019)\citenamefont {Chacko},
  \citenamefont {Du},\ and\ \citenamefont {Geller}}]{Chacko:2018uke}%
  \BibitemOpen
  \bibfield  {author} {\bibinfo {author} {\bibfnamefont {Z.}~\bibnamefont
  {Chacko}}, \bibinfo {author} {\bibfnamefont {P.}~\bibnamefont {Du}}, \ and\
  \bibinfo {author} {\bibfnamefont {M.}~\bibnamefont {Geller}},\ }\href
  {\doibase 10.1103/PhysRevD.100.015050} {\bibfield  {journal} {\bibinfo
  {journal} {Phys. Rev. D}\ }\textbf {\bibinfo {volume} {100}},\ \bibinfo
  {pages} {015050} (\bibinfo {year} {2019})},\ \Eprint
  {http://arxiv.org/abs/1812.11154} {arXiv:1812.11154 [hep-ph]} \BibitemShut
  {NoStop}%
\end{thebibliography}%
\end{document}